\documentclass{optica-article}

\journal{opticajournal} 

\articletype{Research Article}
\usepackage[T1]{fontenc}
\usepackage[latin9]{inputenc}
\usepackage{babel}
\usepackage[normalem]{ulem}
\usepackage[toc,page]{appendix}
\usepackage{doi}
\usepackage{hyperref}
\usepackage{latexsym}
\usepackage{algpseudocode}
\usepackage{mathrsfs}
\usepackage{color,verbatim}
\usepackage{psfrag}
\usepackage{rotating} 
\usepackage{bbold} 
\usepackage{multirow}

\usepackage{colortbl} 

\usepackage{subcaption} 

\usepackage{tikz}

\usepackage{algorithm}

\newcommand{\bra}[1]{\langle#1 |}
\newcommand{\ket}[1]{|#1 \rangle}
\newcommand{\bigket}[1]{\Bigl \lvert#1  \Bigr \rangle}

\newcommand{\average}[1]{\langle #1 \rangle}
\newcommand{\sandwich}[3]{\left \langle #1 \middle \vert #2 \middle \vert #3 \right\rangle}

\definecolor{myDarkGreen}{rgb}{0, 0.7, 0}

\begin{document}

\title{Explicitly Quantum-parallel Computation by Displacements}
\author{ Uchenna Chukwu\authormark{1}, Mohammad-Ali Miri\authormark{1}, Nicholas Chancellor\authormark{1,*}}
\address{\authormark{1}Quantum Computing Inc (QCi), 5 Marine View Plaza Hoboken NJ 07030 USA \\ \today}

\email{\authormark{*}nchancellor@quantumcomputinginc.com}


\abstract{
 We introduce an encoding of information in the relative displacement or photon number of different optical modes. Since the loss rate to interference is insensitive to squeezing and many non-Gaussian fluctuations, such a space is relatively protected from imperfections. We show that photon subtraction protocols can be used to create high-quality quantum superpositions of squeezed states with much higher fidelity than when the protocol is restricted to producing only cat states (superpositions of coherent states).  We also show that the amount of squeezing and anti-squeezing introduced is moderate,  and unlikely to dominate the photon number. This parallel processing allows for explicit use of non-Gaussian interference as opposed to the more incidental role played by non-Gaussianity in all-optical coherent Ising machines. A key observation we make is that displacements of optical states provide a convenient degree of freedom to encode information for quantum parallel processing. Furthermore, we discuss important considerations for realizing an optical quantum annealer based on differential photon number encoding. In particular, we discuss the need to perform quantum erasure on loss channels from interference, as well as the ability to correct degrees of freedom not used for the encoding without disrupting the processed quantum information.
}

\section{Introduction}

Light provides an appealing platform for quantum computation; not only does it propagate freely, enabling high interactivity between information encoded in photons, but it also exhibits weak interactions with its environment. This has led to proposals for gate-based model quantum computing with photonics \cite{Knill2001KLM}.An alternative direction lies in analog optimization, which may provide computational benefits even in the pre-fault-tolerant era of quantum technologies. Quantum annealers, which can be realized in matter-based systems such as flux-qubits or Rydberg atoms \cite{Johnson2011FluxQubitAnnealer,Goswami2024RydbergQA,deOliveira2025RydbergQA} are one example of such analog quantum computing systems.

In the optical setting, coherent Ising machines \cite{wang2013coherent, Yamamoto2017CoherentIsing} present one effort in this direction, but there are questions as to exactly how much they use the underlying quantum effects. Unlike quantum annealing, where a slight generalization of the theoretical model is a universal model for quantum computing \cite{Kempe2006HamComplex,Biamonte2008AQCUniversal} and the benefits of quantum effects are well understood 
\cite{Roland2002,Hastings2013topObstruct,andriyash2017quantummontecarlosimulate,Morley19a}, the underlying theoretical model of coherent Ising machines is much less well understood. Another non-universal quantum paradigm is Gaussian Boson Sampling, which has been adapted to solve optimization problems in a variational setting \cite{bradler2021ORCA, goldsmith2024ORCA_TSP}. 
However, these approaches do not involve directly encoding the problem into the quantum hardware. They rather rely on the ability of the complex distributions produced in the sampling to act as a tool in sampling from optimal regions of the solution space. Other paradigms are emerging such as the entropy computing paradigm \cite{nguyen2024entropycomputing}. 

In this work, we investigate a new approach for encoding information for a photonic quantum annealer. We show that displacement, which roughly corresponds to the creation of classical electromagnetic waves, is a natural way to encode and process information without interference from other degrees of freedom. We argue that the interference between these degrees of freedom provide a natural way to engineer interactions which are insulated from a wide range of other degrees of freedom. We find that we can effectively encode a bit of information in the relative displacement (which in turn relates to photon number) of two packets of fixed phase, with the variable value determined by which mode has more photons.  We call this scheme Differential Photon Number Encoding (DiPNE), and a closely related method based on measuring Differential \emph{Displacement} (DiDE). This is similar to the encoding scheme used in entropy computing. We argue that similar to the coherent Ising machine setting \cite{Leleu2019CIMampCorr}, degrees of freedom which are not directly used to encode the variables (the sum of the displacements and the magnitude of the difference) can be measured without measuring out the encoded information and causing decoherence. Such measurements would allow these degrees of freedom to be corrected. The encoding we explore is effectively the one which is created (briefly) before homodyne measurement is performed in the coherent Ising setting.  

We also explore how non-Gaussian superpositions can be created, overcoming a key weakness of the coherent Ising setting, where the sources of non-Gaussianity are somewhat incidental, coming from saturation of the nonlinear gain material or photon loss \cite{Yamamoto2017CoherentIsing,Yamamura2017quantumIsing}. Such an incidental nature raises questions as to whether the essential elements of coherent Ising machine computation can be captured by classically efficient Gaussian approximations \cite{Clements2017GaussIsing,Tiunov2019simIsing,Tatsumura2021simBif,Ng2022GaussIsing}. This would remove the need for complex optical hardware, but also the potential for quantum advantage. Recent work has shown that coherent Ising machines seem to perform better in the regime with few photons \cite{Kumagai2025singlePhotonCIM}, similar to the one studied here, but it is unclear what role non-Gaussianity plays in this application. 

We discuss how quantum superpostions of displacement-encoded states can naturally be created using a combination of a photon subtraction process and a beamsplitting process inspired by homodyne measurement. This is an extension of protocols to produce Schr{\"o}dinger kitten states \cite{Walschaers2021nonGauss}. We further argue that since our encoding is insensitive to quadrature squeezing by construction, squeezing can also be used to drive superpositions and to homogenize between different photon number measurement results in the subtraction scheme. Effectively, we argue that the encoding scheme we present here is able to combine some of the advantages of Gaussian Boson sampling with those of coherent Ising machines. Namely, it explicitly uses non-Gaussianity in a direct way, as Gaussian Boson sampling does, while having a direct and explicit encoding of an optimization problem through interference effects, like coherent Ising machines. In particular, we discuss a route toward much more direct quantum parallelism, an aim of coherent Ising machines, through a preparation process which induces a high-quality non-Gaussian superposition of squeezed states that are displaced from zero quadrature. This allows an effective encoding in relative photon number. 

In section \ref{sec:scope_and_lim} we lay out the scope and limitations of this work, to provide proper context. Next, in section \ref{sec:enc_num} we introduce the encoding which will be the subject of the work. Next in section \ref{sec:interactions} we discuss how interactions can be implemented in the encoding and their insensitivity to squeezing degrees of freedom. This is show to extend to many non-Gaussian degrees of freedom as well, as discussed in appendix \ref{app:int_gen}. This section also includes brief discussion of some key considerations, such as the need for quantum erasure (section \ref{sub:quant_erase}) and the ability to correct for undesired fluctuations in degrees of freedom not used to encode information without disturbing the information being processed (section \ref{sub:meas_for_corr}). After that in section \ref{sec:ng_drive} we discuss how non-Gaussianity can be introduced and quantum parallel processing could be performed. We first review the idea of Sch{\"o}dinger kitten states (section \ref{sub:schro_kit}) and discuss how they can in principle photon subtraction could be used to generate superposition in our amplitude based encoding scheme (section \ref{sub:enc_sup}), and finally how Gaussian operations can be used to compensate for the non-deterministic nature of photon number measurements (section \ref{sub:Gauss_drive}). Finally in section \ref{sec:num_meth} we review the numerical methods for reproducibility. Finally In section \ref{sec:disc_conc} we discuss and conclude the work. To make this manuscript more self-contained we have included an extended discussion of the background and a review of existing paradigms as appendix \ref{app:background}.

\section{Scope and Limitations\label{sec:scope_and_lim}}

In this work we do not propose a full paradigm for building an analog optical quantum optimiser, rather we we discuss the potential opportunities and some of the challenges for encoding in these systems (and by extensions other systems which can be described as quantized harmonic oscillators).  This is intentional as it is important to understand the encoding in detail before developing a full paradigm. This paper studies one aspect of early entropy computers, the encoding. We emphasize that this work does not fully describe the entropy computing paradigm, which will be discussed in forthcoming work. It is also likely that analogs to what we propose here can also be done in the quadrature encoding setting used in coherent Ising machines, but that is not the focus of the current work. Due to some paralles this work gives some important context on what considerations would be needed to be taken into account if one wanted to extend the coherent Ising paradigm into a bona-fide quantum paradigm, on par with quantum annealing. 

It is worth noting that throughout this work we ignore the decohering effect of photon loss, which in practice could have a major effect on performance. The goal of this work is to mathematically understand what could be possible setting rather than propose a full experimental paradigm, where such analysis would be crucial to making a case for practicality. The aim of this work is to build foundations for developing full paradigms in the future, and as such we do not discuss important real considerations in actually implementing what we have proposed here, possibly the most important of which being the decohering effect of photon loss.

\section{Differential Photon Number Encoding \label{sec:enc_num}}

When considering classical electromagnetic waves, and by extension coherent states of light, there are two complementary degrees of freedom to consider, the amplitude of the wave (mathematically the magnitude of the displacement) and its phase. Coherent Ising machines operate by encoding a binary variable into the phase of the light pulses, using a degenerate optical parametric oscillator to stabilize two possible phases where the nodes of the newly generated waves align with the nodes of the pump light at half the wavelength. To faithfully realize an Ising model for example using optical pulses in the time domain, the displacements of the pulses of light within a coherent Ising machine should be roughly equal, and the inhomogeneity in the amplitudes should be suppressed.

\begin{figure}[h!]
    \centering
        \includegraphics[height=5 cm]{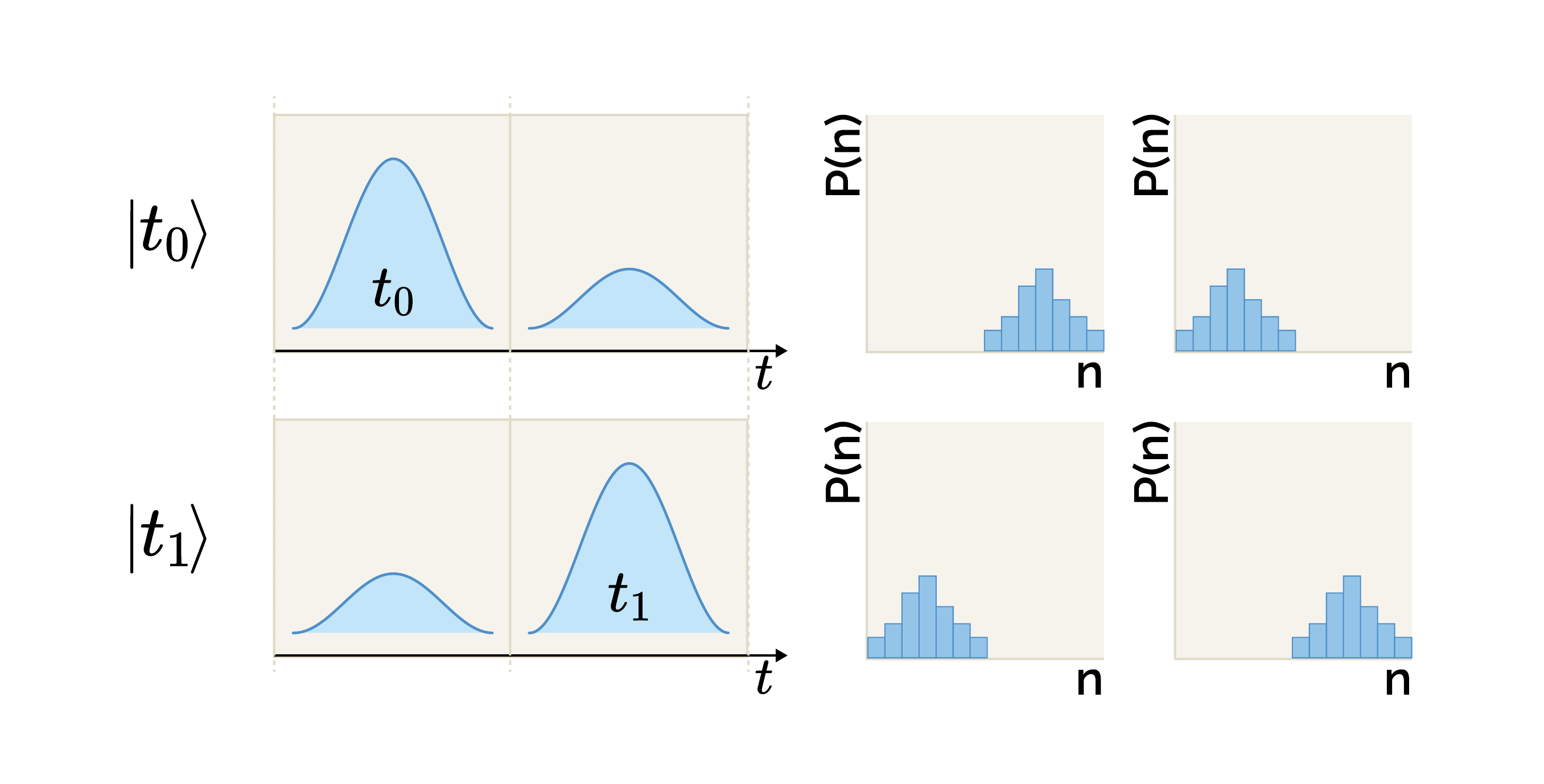}
    \caption{An illustration of DiPNE and DiDE encoding schemes. Both schemes use relative levels of light in two modes, represented as time bins here on the left. The two differ in what is measured DiDE (differential displacement encoding) is based on directly measuring field strength (mathematically represented by displacement). DiPNE (differential photon number encoding) is instead based on photon number measurements, drawn from distributions like those shown on the right.}
    \label{fig:schematic}
\end{figure}

An interesting converse to the coherent Ising machine encoding is to instead encode into displacement magnitudes (or photon number), and to rather fix the phase for controllable interactions. This is effectively the strategy employed in early implementations of entropy computing \cite{nguyen2024entropycomputing}, with occupied bins corresponding to variable values in a time bin encoding. For the current discussion, we will generalize the concept slightly, and instead of considering encoding completely in the presence or absence of light, we will rather consider encoding in the relative displacement magnitudes (or relatedly, relative photon number) in different modes. Two encoding possibilities are represented schematically in figure \ref{fig:schematic}.

Relative displacements of light pulses are what is actually measured in the homodyne measurement procedure \cite{Gerry_Knight_2004}, which is used in coherent Ising machines. In this procedure, a reference pulse (often called the local oscillator) is simultaneously used to illuminate a $50:50$ beamsplitter, along with a pulse to be measured. Interference between the two pulses determines the distribution of light in the output channels of a homodyne measurement device. Reflection causes a phase shift of $+i$ and transmission causes no phase shift. If the pulse being measured has a relative phase of $+i$ compared to the reference pulse, then constructive interference will occur when the reference pulse is reflected and the measured pulse is transmitted, leading to larger displacement on the side which the reference pulse comes in. On the other hand, a relative phase of $-i$ will lead to constructive interference when the measured pulse is reflected and the reference pulse is transmitted. A real phase difference leads to equal displacement in both outputs. 

This can be seen mathematically by considering an initial pair of coherent states acted on by a $50: 50$ beamsplitter

\begin{align}
\ket{\alpha_m}\ket{\alpha_\mathrm{LO}}  \underset{\mathrm{bs}}{\rightarrow} \bigket{\frac{\alpha_m+i\alpha_\mathrm{LO}}{\sqrt{2}}}\bigket{\frac{\alpha_\mathrm{LO}+i\alpha_m}{\sqrt{2}}} \label{eq:homodyne}
\end{align}

\begin{figure}[h!]
    \centering
     \begin{subfigure}{0.48\linewidth}
        \centering
        \includegraphics[width=\linewidth]{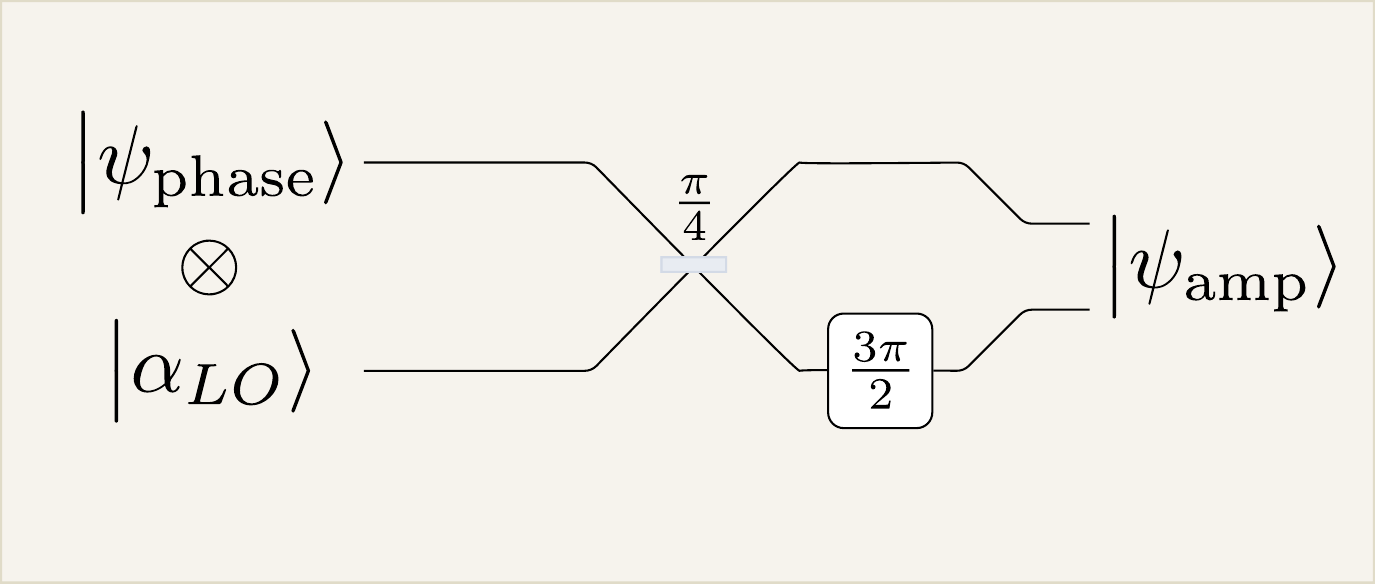}
        \caption{phase to DiPNE/DiDE}
        \label{fig:kit_to_sup}
    \end{subfigure}
    \hfill
    \begin{subfigure}{0.48\linewidth}
        \centering
        \includegraphics[width=\linewidth]{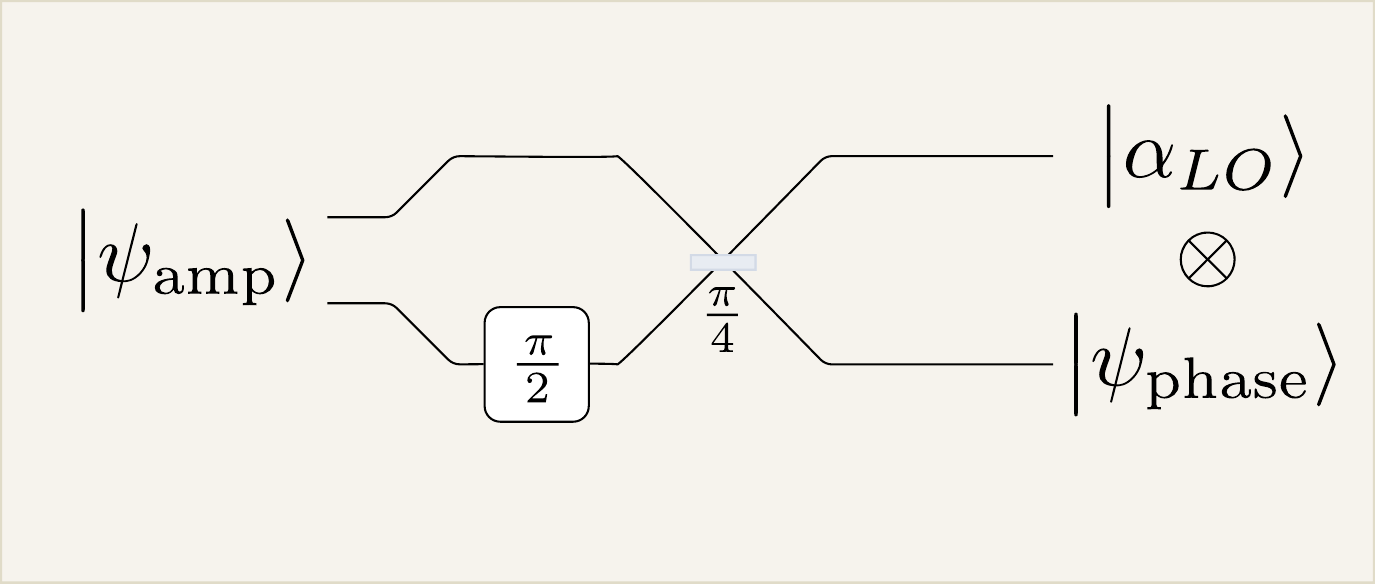}
        \caption{DiPNE/DiDE to phase}
        \label{fig:sup_to_kit}
    \end{subfigure}
    \caption{A schematic of translation (a) between phase representation (as used in a coherent Ising machine) and DiPNE/DiDE representaion. Subfigure (b) shows the reverse translation. Edges entering and leaving rectangles indicate beamsplitting with a given angle, and squares indicate phase shifting. For all schematics in this manuscript the input appears at the left and the output at the right. Note that the phase encoding is confined to a single photon and therefore in a product state with the local operator, whereas the DiPNE/DiDE encoding method based on amplitudes cannot generally be decomposed as a tensor product.}
    \label{fig:kit_to_sup_to_kit}
\end{figure}

Within the context of homodyne measurement, the outputs of the beamsplitter are then measured immediately to determine the relative phase. However, they do not need to be, if we consider light with a relative $\pm i$ phase to the local oscillator (or in superpositions of these, as we will discuss later); these will be translated to encoding in the relative displacement of light in two pulses. As long as $\left|\alpha_\mathrm{LO}\right| > \left|\alpha_m\right|$, the relative phase is fixed, and the two modes can be made to be in phase by applying a $-i$ phase shift to the first mode. This process is shown in figure \ref{fig:kit_to_sup}, while the reverse process which is important for later discussions is shown in figure \ref{fig:sup_to_kit}. Since the unitary operation that a beamsplitter performs is self-inverse, a translation in the other direction is also possible. Relative amplitude information stored in two in-phase pulses can be converted to quadrature information by first applying a $+i$ phase shift to the first mode, applying a $50: 50$ beamsplitter and then discarding the second mode (which would correspond to the local oscillator in homodyne measurement, and only carries information about the total displacement).  

Since average photon number of a coherent state is
\begin{equation}
\bar{n}=\left|\alpha\right|^2
\end{equation}
and the variation in photon number goes as,
\begin{equation}
\Delta n=\bar{n}^\frac{1}{2}.
\end{equation}
It follows from these equations and equation \ref{eq:homodyne} that the relative photon number in each mode will reliably be able to distinguish the phase of the input when (labeling the final modes as $0$ and $1$ respectively)
\begin{align}
\left|\bar{n}_0-\bar{n}_1\right| \gtrapprox \sqrt{(\Delta n_0)^2+(\Delta n_1)^2} \\
\left|\left|\alpha_0\right|^2-\left|\alpha_1\right|^2\right| \gtrapprox \sqrt{\left|\alpha_0\right|^2+\left|\alpha_1\right|^2}.
\end{align}
Assuming that $\alpha_m$ and $\alpha_\mathrm{LO}$ are initially out of phase by $\pm i$, this translates to
\begin{equation}
\left(\left|\alpha_\mathrm{LO}\right|+\left|\alpha_m\right|\right)^2-\left(\left|\alpha_\mathrm{LO}\right|-\left|\alpha_m\right|\right)^2 \gtrapprox \sqrt{\left(\left|\alpha_\mathrm{LO}\right|+\left|\alpha_m\right|\right)^2+\left(\left|\alpha_\mathrm{LO}\right|-\left|\alpha_m\right|\right)^2}.
\end{equation}
Of particular interest here is the case where $\left|\alpha_\mathrm{LO}\right|=\left|\alpha_m\right|$, in which case, one of the output modes of the beamsplitter will be vacuum and this condition collapses to the condition of distinguishability of a coherent state from vacuum by photon counting. It is worth noting that the dual-rail encoding which is used in protocols such as the famous KLM protocol \cite{Knill2001KLM} can be viewed as a special case of DiPNE encoding, where a number state (with a single photon) is used in place of a coherent state allowing for complete distinguishability.

We have shown a relationship between the phase encoding with fixed displacement used in coherent Ising machines and the encoding we explore here called DiPNE (differential photon number encoding). In other words, the explicit encoding of a binary variable $b$ into two modes as
\begin{equation}
b=\begin{cases}
0 & n_0>n_1 \\
1 & n_1>n_0 \\
\mathrm{undefined} & n_0=n_1
\end{cases}. \label{eq:bin_encode_photon}
\end{equation}

Thus far, we have only discussed how coherent states can be used to store and process information, which in itself, inherently, cannot support a quantum advantage, even with single-photon measurement (even Gaussian Boson sampling requires squeezing to become classically intractable). We will, however, later show how highly non-Gaussian states can be naturally generated and used to effectively perform quantum parallel processing in section \ref{sec:ng_drive}. An advantage of representing information in relative photon number as opposed to relative phase is that it is easy to extend beyond binary variables, for example, by extending equation \ref{eq:bin_encode_photon} to more than two modes, where the mode with the most photons corresponds to the variable value. 

A related but distinct encoding is to rather encode a binary variable into relative \emph{displacement}, which we will call differential displacement encoding (DiDE), which can be measured using homodyne measurement \cite{footnote1}. 
\begin{equation}
b=\begin{cases}
0 & \overline{X}_0>\overline{X}_1 \\
1 & \overline{X}_1>\overline{X}_0 \\
\mathrm{undefined} & \overline{X}_0=\overline{X}_1
\end{cases}. \label{eq:bin_encode_disp}
\end{equation}
For coherent states, which are displaced in the $+X$ direction, the definitions of \ref{eq:bin_encode_photon} and \ref{eq:bin_encode_disp} will be equivalent, at least in the average sense, where more displacement leads to more photon counts on average. Where the equivalence breaks down is for more general states of light. Even when considering just Gaussian states, different levels of squeezing on the different modes can shift the relative photon number without changing the displacement. As we discuss later, the squeezing degrees of freedom may be harder to control when performing calculations in superposition, so the DiDE may be favoured in cases with strong and non-uniform squeezing on different modes. In particular, squeezed states have photon number distributions governed by $\bar{n} = |\alpha|^2 + sinh^2(r)$, $\Delta n = \sqrt{|\alpha|^2 e^{2r} + sinh^2(r)}$, where $r$ is the magnitude of the squeezing parameter. One important note is that the outcome of homodyne measurements can be efficiently predicted for an entirely Gaussian system, while photon number measurements cannot (this is the essential idea behind Boson sampling). This implies that when the DiDE encoding in \ref{eq:bin_encode_disp} is used, Gaussianity must be broken in another way. We discuss one way that has a natural interpretation in terms of quantum parallelism in section \ref{sec:ng_drive}.

\section{Interactions\label{sec:interactions}}

We have discussed DiPNE and DiDE encodings which are effectively the converse of the encoding used in coherent Ising systems (fixed phase with information encoded in relative displacement/photon number). Given this encoding, it is worth discussing how optimization problems could be realized for information encoded in this way. In particular, we wish to represent the problem in a similar way to a coherent Ising machine, where the problem is represented in the loss rates, with more favorable states being less prone to loss.
\begin{figure}[h!]
    \centering
        \includegraphics[width=12 cm]{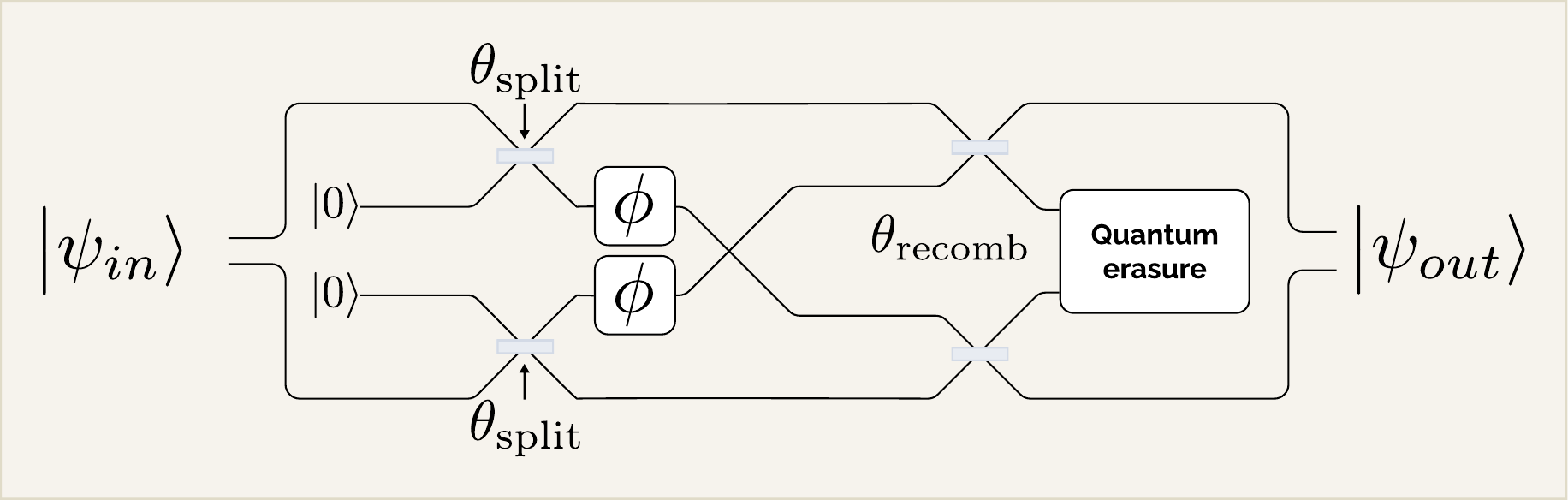}
    \caption{A schematic of how to perform interference between two modes with a photon number or displacement encoding. Edges entering and leaving rectangles indicate beamsplitting with a given angle, and squares indicate phase shifting. To prevent decoherence, quantum erasure must be performed on the exiting light as discussed in section \ref{sub:quant_erase}. For all schematics in this manuscript the input appears at the left and the output at the right.}
    \label{fig:gadget_1}
\end{figure}

Fortunately, if all modes are in phase with each other, it is relatively easy to implement controlled constructive or destructive interference between pulses of light. Since energy is conserved, destructive interference one side of the beamsplitter necessarily implies constructive on the other and vice-versa. To avoid confusion, we will always reference the interference which is happening to the mode which remains within the system, rather than the one sent to erasure As is done in the coherent Ising setting, one can use a beamsplitter to pick off a small portion of the light in a pulse, and then interfere this light with another pulse using a beamsplitter. There will be an initial $+i$ phase shift from the initial beamsplitter, and therefore destructive interference will be present on the mode which remains in the system on the second beamsplitter, meaning that more loss will occur if both modes are occupied This can be done symmetrically between the two modes, as depicted in figure \ref{fig:gadget_1}. Likewise, if a $\pi$ phase shift is added to the light, which is picked off, then the interaction will favor both pulses having a large displacement, as destructive interference on the channel which does not go to erasure will occur. 

In the case of coherent states, the number of photons lost can be calculated analytically. In this case, we start with two modes $\ket{\alpha_1}$ and $\ket{\alpha_2}$. We initially split off some light from each mode using a beamsplitter defined by an angle $\theta_\mathrm{split}$ and then the split off light (taken from mode $1$ or $2$) is then 
\begin{equation}
\ket{i \sin(\theta_\mathrm{split})\alpha_{1,2}}.
\end{equation}
We can then either apply a $\pi$ phase shift to this light or not and perform beamsplitting again between the split off light and the other mode using an angle $\theta_i$. The light exiting the system will therefore be in the states
\begin{equation}
\ket{i\sin(\theta_i)\alpha_{1,2}\pm i \sin(\theta_\mathrm{split})\alpha_{2,1}}.
\end{equation}
The total number of photons leaving the system in each mode is equal to the sum of the square of the magnitude displacement in the two exiting modes; so, therefore, we have
\begin{align}
\bar{n}_\mathrm{tot}=\left|\sin(\theta_i)\cos(\theta_\mathrm{split})\alpha_1\pm  \sin(\theta_\mathrm{split})\cos(\theta_i)\alpha_2\right|^2+ \nonumber \\ \left|\sin(\theta_\mathrm{split})\cos(\theta_i)\alpha_1\pm  \sin(\theta_i)\cos(\theta_\mathrm{split})\alpha_2\right|^2=\\
\left(\sin^2(\theta_i)\cos^2(\theta_\mathrm{split})+\sin^2(\theta_\mathrm{split})\cos^2(\theta_i)\right)\left(|\alpha_1|^2+|\alpha_2|^2\right)\pm \nonumber \\ 
4\sin(\theta_\mathrm{split})\sin(\theta_i)\mathrm{Re}[\alpha_1\alpha^\star_2].
\end{align}
We can compare this to the number of photons which would be lost in the case where no interference is preformed and the split off photons are simply allowed to exit the system,
\begin{align}
\bar{n}_\mathrm{base}=\left(\sin^2(\theta_i)\cos^2(\theta_\mathrm{split})+\sin^2(\theta_\mathrm{split})\cos^2(\theta_i)\right)\left(|\alpha_1|^2+|\alpha_2|^2\right).
\end{align}
The total number of photons lost is therefore
\begin{align}
\bar{n}_\mathrm{tot}=\bar{n}_\mathrm{base}+\bar{n}_\mathrm{interfere} \\
\bar{n}_\mathrm{interfere}=\pm 4\sin(\theta_\mathrm{split})\cos(\theta_\mathrm{split})\sin(\theta_i)\cos(\theta_i)\mathrm{Re}[\alpha_1\alpha^\star_2],
\end{align}
assuming that $\alpha_1$ and $\alpha_2$ are in phase with each other and that $\theta_\mathrm{split},\theta_i <\pi/2$. The interference will then favor (in terms of the overall loss rate) less equal superpositions (effectively anti-ferromagnetic coupling) in the $+$ case and more equal superpositions in the $-$ case (effectively ferromagnetic coupling). To quantify the effect on the loss due to interference, one can compare the number of photons leaving when both modes are occupied, to the sum of the number leaving in the setting where each mode is set to vacuum.

Furthermore, since the covariance and displacement transform separately for Gaussian states \cite{brask2022gaussianstatesoperations}, the modification of the average loss due to interference will be independent of squeezing. Similar arguments can be made for many types of non-Gaussian distortions of the state which are performed before displacement, as argued mathematically in appendix \ref{app:int_gen}. 

\begin{figure}[h!]
    \centering
    \begin{subfigure}{0.48\linewidth}
        \centering
        \includegraphics[width=\linewidth]{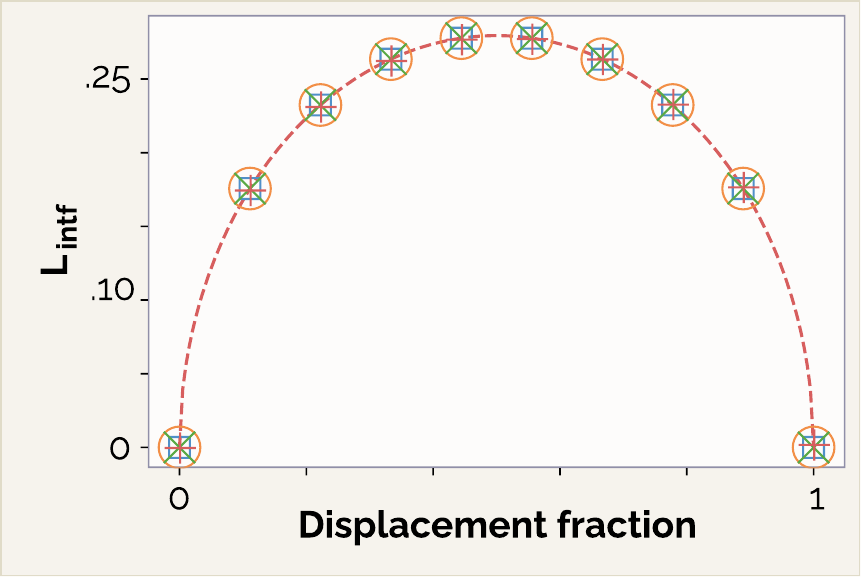}
        \caption{No phase shift}
        \label{fig:interference_loss_af}
    \end{subfigure}
    \hfill
    \begin{subfigure}{0.48\linewidth}
        \centering
        \includegraphics[width=\linewidth]{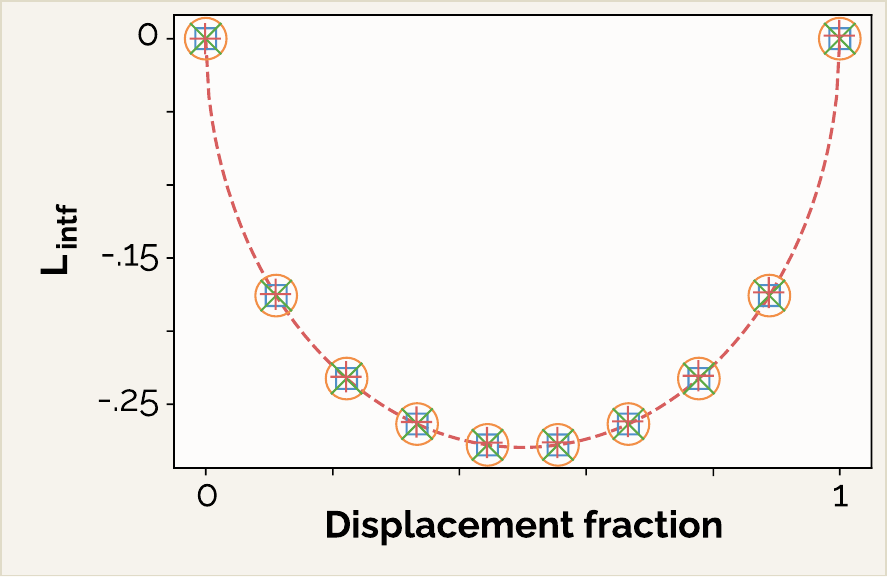}
        \caption{Phase shift of $\pi$}
        \label{fig:inteference_loss_f}
    \end{subfigure}
    \caption{The value of $L_{intf}$ from the gadget shown in \ref{fig:gadget_1} a total of one photon of displacement is added, with the fraction in each mode given on the x-axis. This displacement added to: vacuum in each mode (squares), single photon in mode $0$ and vacuum in channel $1$ (circles), single photon in each mode (diagonal cross), single photon in each with an additional photon of squeezing with an $i$ phase shift in mode $1$ (horizontal cross). No phase shift in (a) and $\pi$ phase shift in (b). The dashed line represents the theoretically calculated value. For both $\theta_\mathrm{split}=\pi/5$ and $\theta_\mathrm{recomb}=\pi/10$. The numerical results were obtained through a full state space simulation with a maximum of $30$ photons in each mode.}
    \label{fig:interference_loss}
\end{figure}

To visualize these effects numerically, we define the following quantity 
\begin{equation}
    L_{intf}=\average{\hat{n}_{int}}_{\ket{\psi_0}\otimes\ket{\psi_1}}-\average{\hat{n}_{int}}_{\ket{\psi_0}\otimes \ket{0}}-\average{\hat{n}_{int}}_{\ket{0}\otimes \ket{\psi_1}}
\end{equation}
where we define 
\begin{equation*}
\average{\hat{n}_{int}}_{\ket{\psi_\mathrm{in}}}
\end{equation*}
as the average number of photons which go to quantum erasure in the scheme depicted in figure \ref{fig:gadget_1}, with $\ket{\psi_\mathrm{in}}$ used as the input to the gadget. Quantum erasure is a process by which information about the logical state of the system is removed from the light leaving the system to prevent decoherence. A discussion of how this can be accomplished in this specific context can be found in section \ref{sub:quant_erase}. An experimental demonstration of the ability to effectively erase information can be found in the induced coherence experiments reported in \cite{Zou1991Induced}. 
This quantity represents the difference between the number of photons lost to the two interaction modes and the sum of the loss when one of the two modes is empty. As a numerical example, we consider this quantity when a single photon of displacement is shared between the two modes. We add a variety of non-Gaussian fluctuations, firstly a single photon added to mode $0$, then a single photon added to both modes. Finally, a single photon added to each as well as one photon each of squeezing performed with a relative factor of $i$ to produce non-Gaussian displaced states. Figure \ref{fig:interference_loss} confirms that all of these scenarios yield the same result, which is in line with theoretical predictions.

One complication which is not present in the coherent Ising setting is the fact that since the value is represented by relative displacements of different modes, the loss from picking off light for interference needs to be compensated in the other mode. Fortunately, this can be achieved in a conceptually simple way by also taking the same amount of light from the pulse which does not participate in the interaction. An intentional imbalance can be used to encode single-body terms.

A more subtle issue comes if fully quantum operation is desired. In particular, to process information in superposition, coherence needs to be maintained. This means that leakage of information about the state of the quantum system into the outside world should be minimized. This is potentially an issue for both the encoding discussed here and the phase-based coherent Ising encoding.  In particular, the energy lost from the channel of the beamsplitter which does not continue will carry information about whether or not the Ising variables agree, in addition to the phase information itself which may be less of a concern as it is only meaningful when compared with the local oscillator.

This complication arises because beamsplitting conserves photon number and information about the values of the variables is contained in the photons leaving the arm of the beamsplitter which does not encode a variable. For example, in coupling, which enforces two variables to not take the same values, more light will exit the system when they do take the same value. If this light is not treated carefully, it could ruin any large-scale coherent superposition taking place in the system. We discuss how this problem can, in principle, be overcome in subsection \ref{sub:quant_erase}.

Furthermore, in the same way, displacement magnitude needs to be controlled to faithfully realize an Ising model in the coherent Ising setting. The combined displacement of the two pulses and the ratio between the more and less occupied one, needs to be controlled in a setting where relative displacement encoding is used. This requirement may be lessened somewhat in settings where the problem does not need to be represented at a high precision. For example, maximum independent set is an NP-hard problem where strong interactions are needed to enforce that nodes do not share an edge, but these do not need to be carefully controlled to faithfully represent the problem. Even in these cases, some correction may be desired, which means that these quantities need to be measured and feedback needs to be applied. We discuss how these measurements can be done without disturbing the relevant quantum information of the system in section \ref{sub:meas_for_corr}.

\subsection{Quantum Erasure \label{sub:quant_erase}}

 A key idea in quantum information is the concept of quantum erasure. The key idea of quantum erasure is to make said information unavailable, thereby restoring quantum coherence that would otherwise be lost.  These effects have long been demonstrated experimentally. For example, in \cite{Zou1991Induced}, beam paths overlapped to erase information about which path a photon had taken. When the beam paths overlapped, interference was restored, but when a beam block prevented the overlap, this signature vanished, and the erasure was no longer effective.

The question we must ask to be able to implement a fully quantum optical optimizer is whether the information stored in the amplitudes of the light which leaves the beamsplitters can effectively be erased. Theoretically this could be accomplished in a number of ways; for example, if light is mixed and $\pm i$ out-of-phase \cite{footnote2} (or incoherently, without a well defined fixed relative phase) with a much stronger coherent state of light, then the fluctuations in the stronger light can overwhelm the original pulse. Mathematically, this can be seen by observing that in this situation displacements would add in quadrature, and by taking the Taylor expansion of the norm of the sum of their out-of-phase displacement amplitudes, we see that
\begin{equation}
\label{eq:erasure_displacement}
|\alpha| = \sqrt{\left|\alpha_\mathrm{strong} \right|^2+\left|\alpha_\mathrm{weak}\right|^2}=\left|\alpha_\mathrm{strong} \right|+\frac{\left|\alpha_\mathrm{weak}\right|^2}{2 \left|\alpha_\mathrm{strong}\right|}+O(\frac{\left|\alpha_\mathrm{weak}\right|^3}{ \left|\alpha_\mathrm{strong}\right|^2})
\end{equation}
where $|\alpha_{weak}| \ll |\alpha_{strong}|$. The weak component is being 'erased' in amplitude because its contribution to the displacement becomes sub-leading.

If this displacement in Equation \ref{eq:erasure_displacement} were directly measured, the difference between $|\alpha|$ and $|\alpha_{strong}|$ will be suppressed as $\frac{\left|\alpha_\mathrm{weak}\right|}{\left|\alpha_\mathrm{strong}\right|}$ decreases, eventually making the difference unnoticeable. Moreover, the probability of measuring a given number of photons will follow a Poisson distribution 
\begin{equation}
P_n=e^{-\left|\alpha\right|^2}\frac{\left|\alpha\right|^{2n}}{n!},
\end{equation}
Since this is a smooth function of $\left|\alpha\right|$, the difference in the probability of detecting any given number of photons will also vanish as $\frac{\left|\alpha_\mathrm{weak}\right|}{\left|\alpha_\mathrm{strong}\right|}$ vanishes. While there are likely to be more practical ways of erasing information that may leak from the system than mixing with a strong coherent state and performing number or displacement measurements, these simple calculations serve to show that erasure of outgoing photon states is possible as long as the phase is known up to a minus sign (a phase difference of $\pm i$ can be ensured).  Note that because of the fact that it only needs to be known up to a minus sign, these arguments apply equally well to the quadrature encoding used in coherent Ising machines and the encoding we have discussed, where phase is fixed and relative displacement is used to store information. It is also worth observing that the arguments here would also naturally extend to a case where the phase information is completely removed, in other words, where the light is added incoherently to a strong coherent state.

\subsection{Measurements for correction \label{sub:meas_for_corr}}

An additional question is whether it is possible to take measurements of the total displacement and magnitude of the difference between the pulses which encode a binary variable. A hint at how to do this lies in the translation between DiPNE and the phase encoding used in a coherent Ising machine discussed in section \ref{sec:enc_num}. Namely, we translate by using a beamsplitter to take light from each channel, phase shift one of them by $i$, then apply beamsplitting. This is the process depicted in figure \ref{fig:sup_to_kit}.  The encoded variable value (the protected quantum information) would translate to the phase of one of the output channels, while the displacements will correspond to degrees of freedom which need correcting, corresponding to the total number of photons in both channels and the magnitude (but not sign) of their difference. This is similar to a coherent Ising machine, where the amplitude of the pulses can be measured and corrected to ensure that the model is faithfully implemented\cite{Leleu2019CIMampCorr}.

\section{Non-Gaussian driving \label{sec:ng_drive}}

Thus far, our theoretical discussion has mostly focused on processing information with coherent states. However, by themselves, coherent states cannot perform meaningful quantum computations, as they have an efficient classical description. In fact, even Gaussian states, which include the possibility of squeezing, can be efficiently classically described, as long as the set of allowed measurements is limited. 

On the other hand, the behavior of coherent states is convenient to understand mathematically and provides a useful tool for encoding computations because of their easy-to-understand behavior. Because of this, one route to optical computation would be to implement these coherent state operations in quantum superposition. Such processing would be fundamentally non-Gaussian. On the other hand, while an individual coherent state is Gaussian, superpositions of coherent states (for example, a ``Schr{\"o}dinger cat'' state which superimposes coherent states with opposite phase) are highly non-Gaussian. We further show that with DiPNE/DiDE encodings, it is possible to drive such superpositions.

\begin{figure}[h!]
    \centering
    \begin{subfigure}{0.48\linewidth}
        \centering
        \includegraphics[height=2.3 cm]{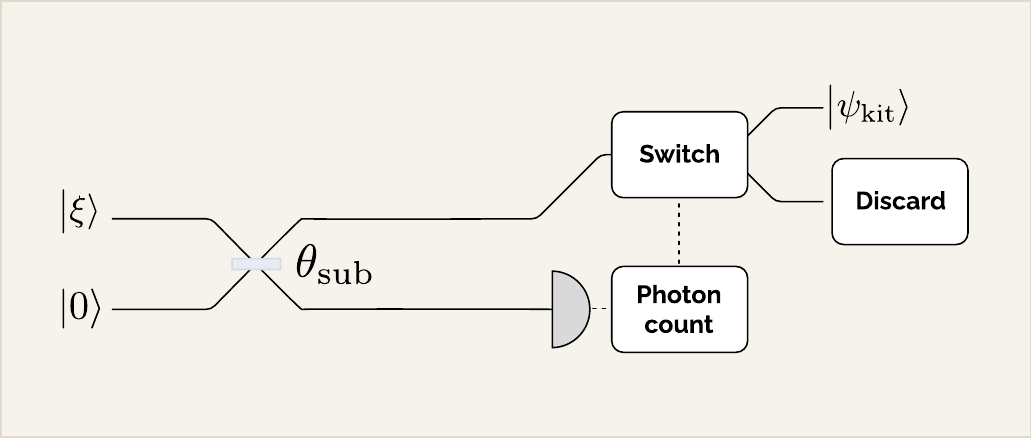}
        \caption{Creation of kitten state}
        \label{fig:kit_create}
    \end{subfigure}
    \begin{subfigure}{0.48\linewidth}
        \centering
        \includegraphics[height=2.3 cm]{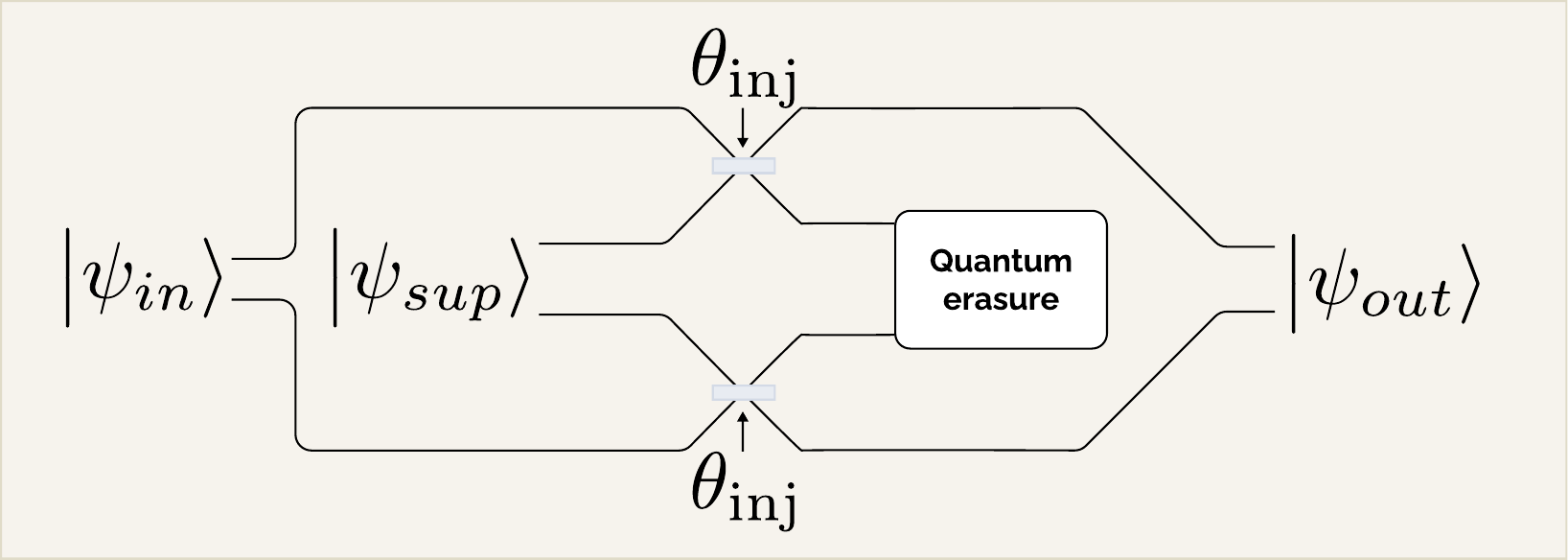}
        \caption{Injection of superposition state}
        \label{fig:kit_inject}
    \end{subfigure}
    \caption{(a) A schematic of how to create a kitten state. Note that this operation only produces kitten states probabilistically, if the correct number of photons is measured. (b) A schematic of an injection scheme after translation to a superposition state using the method depicted in figure \ref{fig:kit_to_sup}.  Edges entering and leaving rectangles indicate beamsplitting with a given angle. Dashed lines in subfigure (a) indicate classical control operations.  For all schematics in this manuscript the input appears at the left and the output at the right.}
    \label{fig:kit_create_inject}
\end{figure}

\subsection{Schr{\"o}dinger Kitten States \label{sub:schro_kit}}

To start with, we consider a well-known quantum state of light known as the Schr{\"o}dinger kitten state \cite{Walschaers2021nonGauss} (hereafter ``kitten state'' for brevity). The kitten state is conceptually similar to the well-known cat state, but can be prepared in a conceptually simple way. A kitten state is a quadrature squeeze vacuum with photons subtracted.

Such a state can, in principle, be prepared by performing beamsplitting on a squeezed vacuum, followed by performing a photon-counting measurement on one output. In the case where a single photon is measured, then the other output must contain a kitten state comprised of odd number Fock states. Schematically, the process for producing such a state is depicted in figure \ref{fig:kit_create} where $\ket{\xi}$ is squeezed vacuum. From this state, we can apply the procedure described in Section \ref{sec:enc_num} (specifically depicted in Figure \ref{fig:kit_to_sup}) to produce an approximate superposition of relative number encoded states. This state can then be injected into the system as a form of non-Gaussian driving, in the matter schematically depicted in figure \ref{fig:kit_inject}.

General arguments can be made that a state prepared by beamsplitting between squeezed and unsqueezed vacuum followed by number measurements will prepare a cat-like state.  Creating approximate superpositions in this way has been studied extensively in other works; see, for example, \cite{Dakna1997CatSubtract,Molmer2006NonGauss,Parigi2007PhotonAddSubtract,Walchaers2017nonGauss,Takase2021CatSubtract,Walschaers2021nonGauss}. While this topic is well studied,  it is still worth reviewing some of the context and providing some additional results. Let us consider a photon subtracted state generated by the procedure depicted in figure \ref{fig:kit_create}.  We consider the case where $k$ photons have been measured; in this case, any state with $n-k\ge 0$ photons in the right channel at the end can only have originated from a state with $n$ photons initially in the right channel, since the left channel is empty. We can therefore efficiently write the coefficients of the state produced by this device as 
\begin{equation}
    C_{n-k}=\frac{1}{\mathcal{N}}\sqrt{\frac{n!}{(n-k)!}}\cos^{n-k}(\theta_\mathrm{sub})C_n(\xi) \label{eq:squeeze_split}
\end{equation}
where 
\begin{equation}
    \ket{\psi}=\sum_{m=0}^\infty C_m \ket{m},
\end{equation}
and
\begin{equation}
    \ket{\xi}=\sum_{n=0}^\infty C_n(\xi) \ket{n}=S(\xi)\ket{0},
\end{equation}
 where $S(\xi)$ is the squeezing operator and $\mathcal{N}$ is a normalisation factor, which is not important for the present discussion. Note that we have absorbed an $n$-independent factor of $\frac{\sqrt{k!}(i)^k}{\sin^{k}(\theta_\mathrm{sub})}$ into this term.
From equation \ref{eq:squeeze_split} we can make several observations. Firstly, the alternating structure of the squeezed state is preserved, since a squeezed state will only have support under even photon numbers $C_{2m+1}(\xi)=0 \text{, where } m\in \mathbb{Z}_+$. Hence, for an even $k$, the final wavefunction will only contain  even photon number states, and for an odd $k$, only odd states. Moreover, the relative phase of the levels will be preserved. These two facts together imply that all states produced in this way will maintain the reflection symmetry of the squeezed state, and will have a much larger second moment in one quadrature direction than the other.

Furthermore, the $n$-dependent prefactor from equation \ref{eq:squeeze_split} has an interesting property. It is comprised of the square root of a polynomial that grows monotonically with $n$ for $n>k$ and contains a maximum power of $n^{k}$ multiplied by a term which is exponential in $n$. For $k>0$, the polynomial term will dominate at low values of $m=n-k$. At high values, the exponential term will dominate. This suggests that these weighting factors will have a single peak at a finite value of $m$, suppressing displacement for both low and high values of $m$. A crude approximation, which is valid for large $n$, is to take only the highest-order term of the polynomial,
\begin{equation}
\sqrt{\frac{n!}{(n-k)!}}\cos^{n-k}(\theta_\mathrm{sub}) =n^\frac{k}{2}\cos^{n-k}(\theta_\mathrm{sub})+O\left(n^{\frac{k-1}{2}}\cos^{n-k}(\theta_\mathrm{sub})\right),
\end{equation}
and then by squaring (since the location of the peak is the same for the square of the coefficient as for the coefficient itself) and setting their derivative with respect to $n$ equal to zero, we can then approximate the value of $n$ at which the peak occurs:
\begin{align}
k\bar{n}^{k-1}\cos^{2\bar{n}-2k}(\theta_\mathrm{sub})\approx-2\bar{n}^k\cos^{2\bar{n}-2k}(\theta_\mathrm{sub})\ln(\cos(\theta_\mathrm{sub})) \\
k\bar{n}^{k-1}\approx-2\bar{n}^k\ln(\cos(\theta_\mathrm{sub})) \\
\bar{n}\approx-\frac{k}{2\ln(\cos(\theta_\mathrm{sub}))} \label{eq:peak_approx}.
\end{align}
This weighting implies that the Husimi Q function (see: \cite{Gerry_Knight_2004}) should peak away from zero, but the (anti-)symmetry imposed by the alternating structure and the phases, implies both that the state should be (anti-)symmetric and that it should have a much larger second moment along one direction than the other. A general class of states which meet these conditions are Schr{\"o}dinger cat-like states, with a positive or negative superposition of two quadrature peaks. We will show numerically that, to a very good fidelity, the states produced by this process look like superpositions of squeezed states. Note that by previous arguments about squeezing not affecting loss, we can accept a superposition of squeezed coherent states rather than demanding a superposition of unsqueezed coherent states. We still want to keep track of the level of squeezing, as too much antisqueezing could make measurement of the displacement difficult in practice, or lead to cases where relative photon number does not match relative displacement.

We can further use this formula to estimate the magnitude of the displacement of the output state by equating the peak value of $n$ to the average number of photons from displacement since $\bar{n}=|\alpha|^2$, which implies that 
\begin{equation}
|\alpha|\approx \sqrt{-\frac{k}{2\ln(\cos(\theta_\mathrm{sub}))}}.
\end{equation}
Effectively, the angle $\theta_{sub}$ can be used to control the level of displacement applied to each state in the cat-like superposition.

We can make a few further observations. Firstly, and perhaps somewhat counterintuitively, according to equation \ref{eq:peak_approx}, the number of photons in the final output state should grow, not shrink, with the value of $k$. This is because more photons being taken away tend to post-select on initial conditions where more photons are present. Recall that the value of $k$ is not chosen as part of the experiment but is a probabilistic measurement value. Another observation we can make here is that, given a measured value of $k$, the final distribution is well defined, even in the presence of very strong squeezing, and even has a well-defined limit when squeezing \cite{Gerry_Knight_2004} is taken to infinity,
\begin{equation}
    C_{2m}(\xi\rightarrow e^{i\theta}\infty) \propto (-1)^m \frac{\sqrt{(2 m)!}}{2^m m!}e^{i\theta m},
\end{equation}
where we have slightly abused the notation in treating  $\infty$ as a positive real number. While the normalization for such a state does not converge (the displacement is only weakly dependent on $m$, $C_{m+1}=-e^{i \theta}\sqrt{\frac{m}{m+1}}C_{m-1}$ \cite{Gerry_Knight_2004}), the prefactors in equation \ref{eq:squeeze_split} will force convergence for a finite $k$. Naively, it would appear that even with arbitrarily strong squeezing, the magnitude of the displacement of the superimposed states is limited by the ability to resolve a high number of photons. However, here we can make an observation, in  \ref{eq:peak_approx}, that the expected photon number also increases as $\theta_\mathrm{sub}$ decreases, since more photons will be post selected if transfer through the beamsplitter is less likely. Mathematically speaking, by balancing beamsplitting angle and squeezing, it is possible to make a kitten state with arbitrarily many photons for a fixed value of $k$.

Furthermore, we can study the quality of the approximation to cat states and generalizations of them, which we can produce by the photon subtraction methods, and while this has been studied elsewhere\cite{Dakna1997CatSubtract,Molmer2006NonGauss,Parigi2007PhotonAddSubtract,Walchaers2017nonGauss,Takase2021CatSubtract,Walschaers2021nonGauss}, we are interested in slightly different properties.  In particular,  since the addition of squeezing does not effect the interference loss,  we are willing to accept states which correspond to a superposition of squeezed and displaced states,  rather than just a superposition of coherent states (the conventional definition of a cat state). For concreteness, these states take the form
\begin{equation}
    \ket{\psi_\mathrm{cat\,sq}}=\frac{1}{\sqrt{2}}\left(D(\alpha)+\exp(i\phi)D(-\alpha)\right)S(\xi)\ket{0}\label{eq:cat_gen}
\end{equation}
where $D(\alpha)$ is a displacement operator and $S(\xi)$ is a squeezing operator.

\begin{figure}[h!]
    \centering
    \begin{subfigure}{0.48\linewidth}
        \centering
        \includegraphics[width=\linewidth]{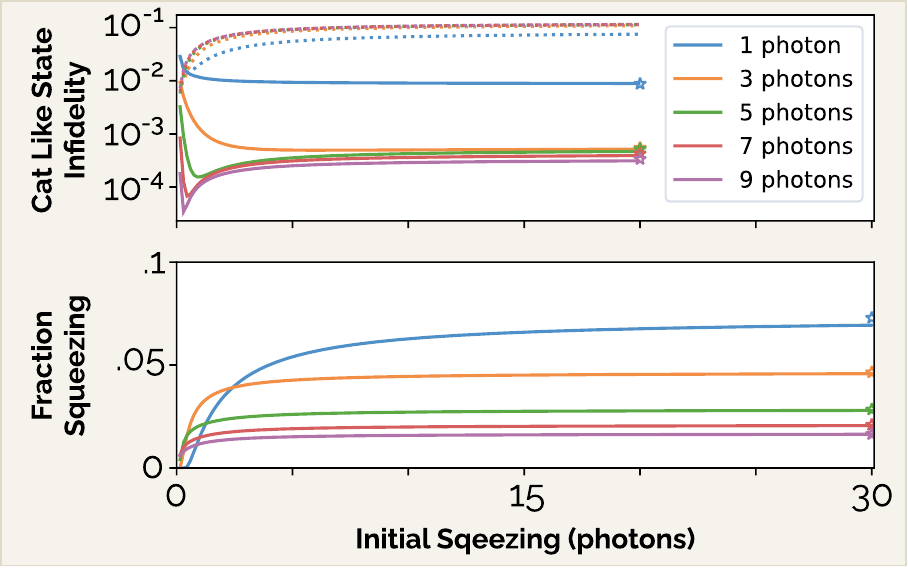}
        \caption{Fit statistics}
        \label{fig:kitten_infid_and_squeezing}
    \end{subfigure}
    \hfill
    \begin{subfigure}{0.48\linewidth}
        \centering
        \includegraphics[width=\linewidth]{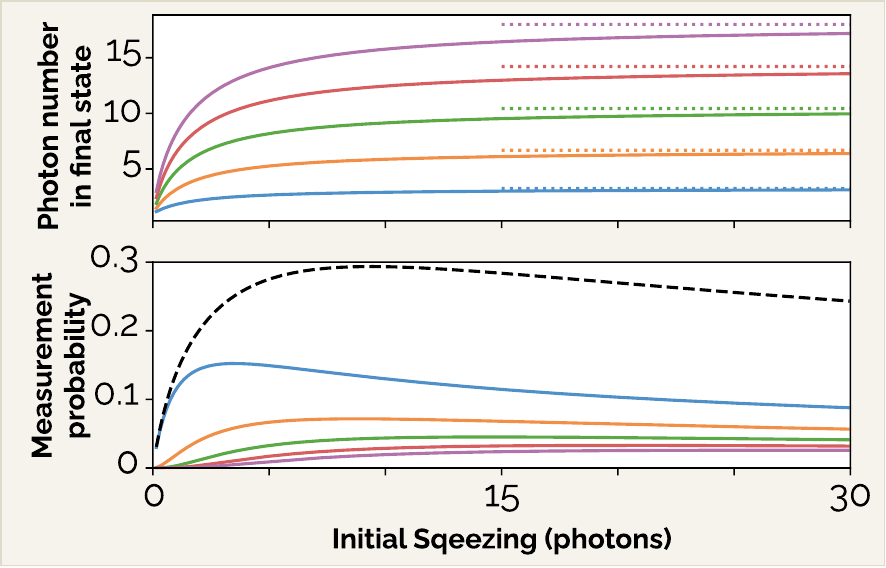}
        \caption{Probability and counts}
        \label{fig:kitten_prob_and_num}
    \end{subfigure}
    \caption{ Numerical results for kitten states produced using the scheme depicted in figure \ref{fig:kit_create} with a fixed angle of $\frac{\pi}{5}$ and a photon count of $k$ for a given number of photons of squeezing in the initial state.  Subfigure (a) depicts the infidelity with a cat-like state (including squeezing in each state in the superpostion as in equation \ref{eq:cat_gen}) and the fraction of photons from squeezing which achieves maximum fidelity.  The stars in this subfigure indicate the infinite squeezing limit. The dotted lines with the same colour coding are infidelity with regular cat states of matching energy (as are more traditionally found in the literature). Subfigure (b) depicts the number of photons and probability of a given measurement device.  The colour coding is the same as in (a), while the dotted lines indicate the infinite squeezing limit.  On the bottom subplot the dashed black line indicates the cumulative value of all probabilities. Calculations were performed in a truncated Hilbert space with a maximum of $1000$ photons in the mode.  Initial squeezed state displacements were calculated using an analytical formula, rather than through operators in the truncated space.}
    \label{fig:kitten_squeeze}
\end{figure}

When fitting against these generalised cat-like states, we fix the total number of photons to be the same number as the kitten state, and then numerically optimize the fidelity with respect to the fraction of photons which come from squeezing. We find in figure \ref{fig:kitten_infid_and_squeezing} that the fidelity is high. For reasonable squeezing, we find that fidelity around $99\%$ are possible for a single photon subtraction; for three or more subtracted photons, a fidelity around $99.9\%$ is possible, with gradual improvement as the number of photons measured is increased. This continues all the way until the infinite squeezing limit.  In contrast, if we demand a true cat state, with no squeezing, the fidelity drops to about $90\%$. We also find that the fraction of photons due to squeezing is relatively small,  less than $10\%$ in all cases and less than $5\%$ if more than one photon is subtracted.  We do not know the origin of the qualitative differences between the single photon subtraction curves and those for subtracting more photons, and this is likely an interesting direction for future study, but is beyond the scope of this paper. 

\begin{figure}[h!]
    \centering
    \begin{subfigure}{0.48\linewidth}
        \centering
        \includegraphics[width=\linewidth]{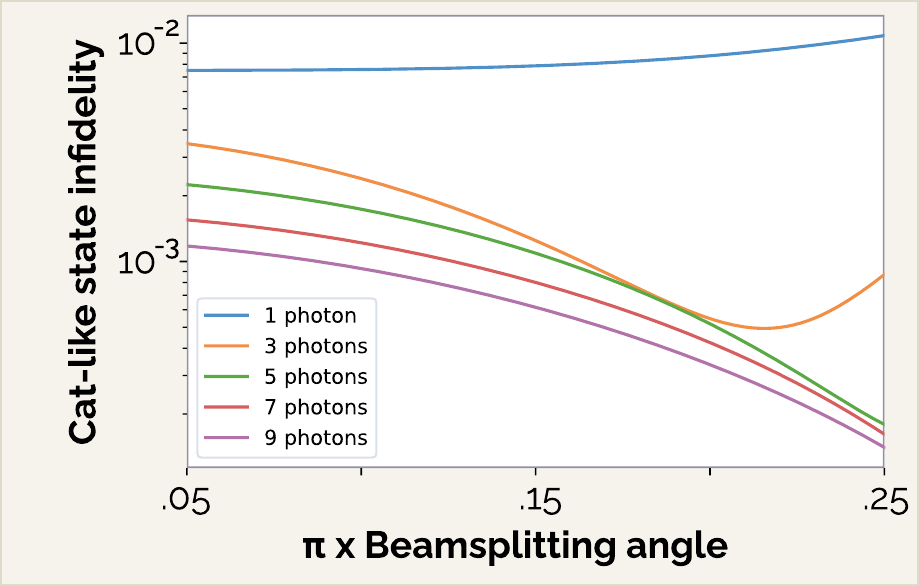}
        \caption{Fidelity}
        \label{fig:fidelity_beamsplit_angle}
    \end{subfigure}
    \begin{subfigure}{0.48\linewidth}
        \centering
        \includegraphics[width=\linewidth]{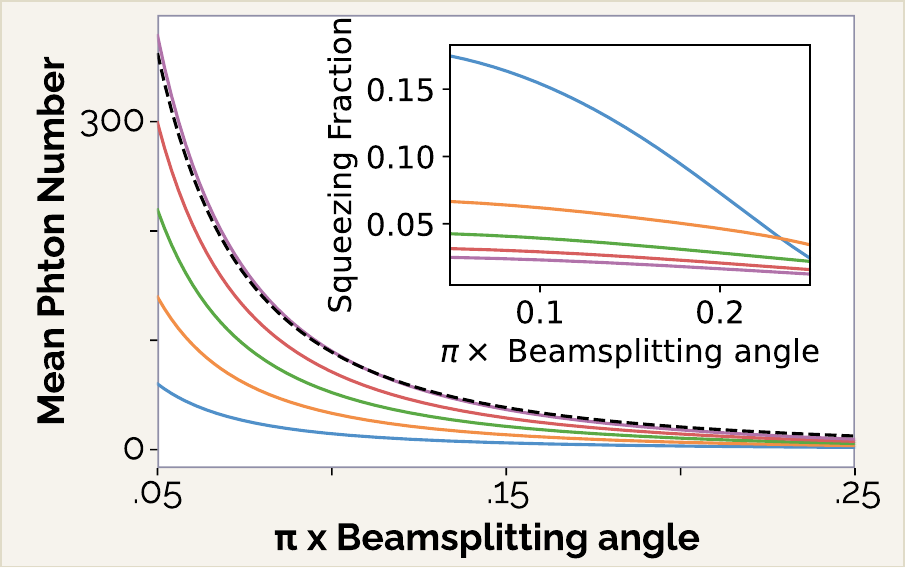}
        \caption{Photon Number}
        \label{fig:photon_number_beamsplit_angle}
    \end{subfigure}
    \caption{Numerical results for kitten states produced using the scheme depicted in figure \ref{fig:kit_create} in the infinite squeezing limit with a photon count of $k$ for variable beamsplitting angle Subfigure (a) depicts the infidelity with a cat-like state (including squeezing in each state in the superposition as in equation \ref{eq:cat_gen}) and the fraction of photons from squeezing which achieves maximum fidelity.  Subfigure (b) depicts the average number of photons in the final state.  The colour coding is the same as in (a), the dashed black line is the peak value expected in the $k=9$ case using the approximation in equation \ref{eq:peak_approx}. The inset of figure (b) depicts the number Calculations were performed in a truncated Hilbert space with a maximum of $1500$ photons in the mode.  Initial squeezed state displacements were calculated using an analytical formula.}
    \label{fig:fidelity_number_beamsplit_angle}
\end{figure}

As figure \ref{fig:kitten_prob_and_num} shows, the likelihood of detecting a specific photon count k after photon subtraction generally decreases as k increases. The cumulative probability of measuring an odd number of photons between $1$ and $9$ inclusive, can be as high as $30\%$ and only drops off slowly with increased squeezing, including into ranges which give results similar to the infinite squeezing limit.  These numerical results agree with theoretical predictions, with the average number of photons increasing linearly with the number of photons detected.  We further can see in figure \ref{fig:fidelity_number_beamsplit_angle} that changing the beamsplitting angle can substantially increase the photon number in the output for a given measurement, while only modestly degrading the fidelity with a superposition of squeezed states and the fraction of photons from squeezing.

\subsection{Encoding Superpositions \label{sub:enc_sup}}

\begin{figure}[h!]
    \centering
    \begin{subfigure}{0.48\linewidth}
        \centering
        \includegraphics[width=\linewidth]{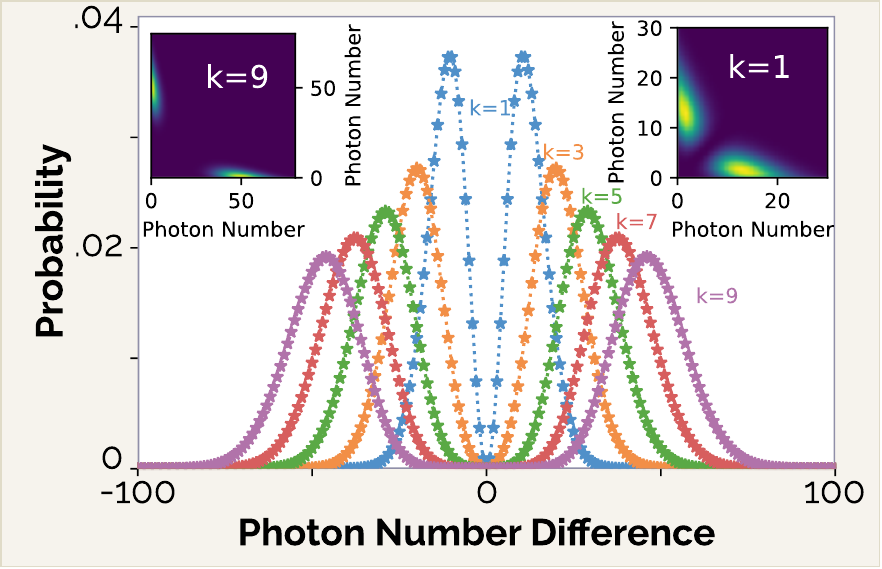}
        \caption{Odd number subtraction}
        \label{fig:number_difference_odd}
    \end{subfigure}
    \hfill
    \begin{subfigure}{0.48\linewidth}
        \centering
        \includegraphics[width=\linewidth]{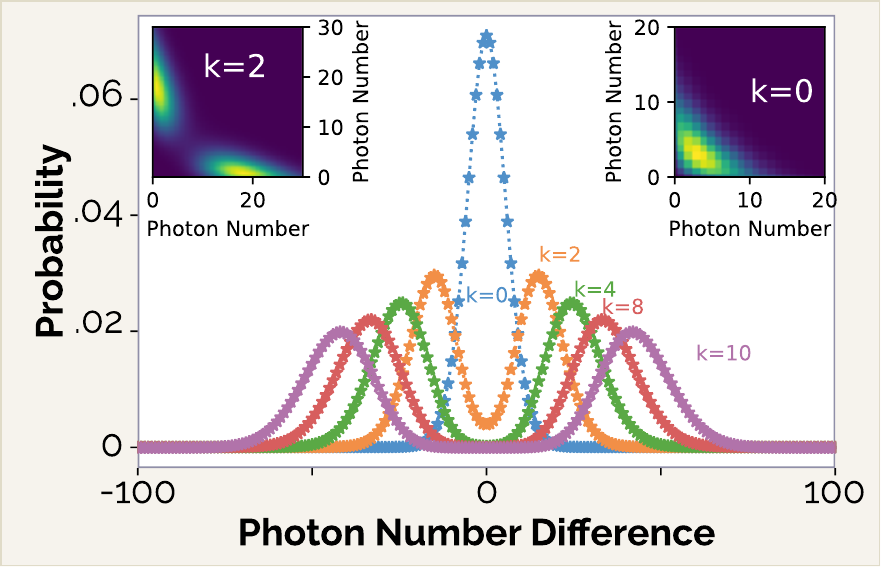}
        \caption{Even number subtraction}
        \label{fig:number_difference_even}
    \end{subfigure}
    \caption{Difference in photon number for kitten state creation as shown in figure \ref{fig:kit_create} followed by beamsplitting as depicted in figure \ref{fig:kit_to_sup}.  For this plot we consider a beamsplitter angle of $\frac{\pi}{5}$ in the kitten state creation, using a squeezed vacuum with $10$ photons of squeezing.  When converting to the amplitude representation, we add a coherent state with a displacement equal to the square root of the number of photons in the kitten state,  plus $2$.  Subplot (a) depicts the case where an odd number $k$ of photons are detected,  and (b) for an even number.  Each curve is labeled with detection number. the insets show heatmaps of the photon number probability distribution for different numbers of photons detected. Calculations were performed in a truncated Hilbert space allowing a maximum of $100$ photons per mode. Squeezed states are initialized using an analytic formula for the occupation.}
    \label{fig:number_difference_odd_even}
\end{figure}

Next, we consider what happens when this state is converted to a displacement representation as depicted in figure \ref{fig:kit_to_sup}. If we consider a true cat state being sent in, then, by the superposition principle,  the two possible quadratures can be considered separately, and the result is another superposition.  Interference of coherent states with equal magnitude displacement and a $\pm i$ phase difference results in all displacements in one output or the other.  For our scheme, we want a well-defined phase difference between the two inputs so we would make $|\alpha_{LO}|$ stronger.  In this case the results is a superposition of states where there are coherent states in both channels but one dominates. This corresponds to a superposition of $0$ and $1$ states as defined in equation \ref{eq:bin_encode_photon} or  \ref{eq:bin_encode_disp}.  We can see in figure \ref{fig:number_difference_odd_even} the photon numbers for different kitten states indeed follows this pattern,  which is unsurprising given the high fidelity of states prepared by photon subtraction from squeezed states. 

One aspect which we have not discussed yet of cat states is that there is a relative phase between the two coherent states; a general cat state is defined as
\begin{equation}
\label{eq:cat_state}
\ket{\psi_\mathrm{cat}}=\frac{1}{\sqrt{2}}\left(\ket{\alpha}+\exp(i \phi) \ket{-\alpha}\right)
\end{equation}
In the case of $\phi=0$ or $\phi=\pi$ the relative phase is $\pm$; only states with an even or odd number of photons are supported.  These are the cases which a kitten state can realise, since squeezed states only contain even numbers of photons, where photon subtraction of an odd (even) number of photons will result in a state with  only odd (even) numbers. For the odd superpositions, translation to the displacement representation lead to an interesting result, which can be seen numerically in figure \ref{fig:number_difference_odd_even}; states with exactly equal number of photons are never produced.  This can also be understood intuitively from superpositions of coherent states. When the number of photons output in each mode is equal, then the contribution from the positive and negative displacements in the cat state is also equal; if they have a relative minus sign, this will lead to cancellation. So, the relative minus sign leads to complete cancellation.

Going beyond the pure cat state picture, we can argue that equal numbers will never be seen from an extension of the famed Hong-Ou-Mandel effect \cite{Hong1987HOM,Gerry_Knight_2004}. Namely, the generalization we show is that if the initial number of photons on both inputs of a $50:50$ are odd, then it is not possible for an equal number of photons to appear on each side of the beamsplitter after. The Hong-Ou-Mandel effect represents the specific case with one photon in each input. This further directly implies that if one input has a superposition of only odd photon numbers, the output numbers cannot be the same following interference with an arbitrary state of light in the other channel. Consider the result of performing a $50:50$ beamsplitting operation on two number states with numbers $n$ and $m$.  This operation can be represented as 
\begin{equation}
\frac{1}{\sqrt{n!m!}}\left(\hat{a}_0^\dagger \right)^n\left(\hat{a}_1^\dagger\right)^m\ket{0} \rightarrow \frac{1}{\sqrt{n!m!2^{n+m}}}\left(\hat{a}_2^\dagger+i\hat{a}_3^\dagger \right)^n\left(\hat{a}_3^\dagger+i\hat{a}_2^\dagger\right)^m\ket{0}.\label{eq:bs5050_op}
\end{equation}
To calculate the probability of having an equal number of output photons, we want to limit ourselves to the cases where the output states take a form proportional to
\begin{equation}
\left(\frac{n+m}{2}!\right)^{-1} \left(\hat{a}_2^\dagger \right)^{\frac{n+m}{2}}\left(\hat{a}_3^\dagger \right)^{\frac{n+m}{2}}\ket{0}.
\end{equation}
We now consider the most and the least photons which can come from each initial mode.  Without loss of generality, let us assume that $n\ge m$.  If all photons in mode 1 are reflected (end in mode 2) then $\frac{n-m}{2}$ photons must be reflected from mode 0 to mode 3.  On the other extreme,  if all photons from mode 1 are reflected to mode 2, then $\frac{m+n}{2}$ photons must be transmitted.  We now need to consider the prefactors on the terms corresponding to each of these.  Since all operators in equation \ref{eq:bs5050_op} commute,  this amounts to counting the terms in a polynomial.  If $k$ photons are reflected from mode $0$ to mode $3$,  these factors can be written as a binomial expression,
\begin{equation}
\binom{n}{k}\binom{m}{k-\frac{n-m}{2}},
\end{equation}
 effectively counting the number of terms each creation operator from each side can come from. Note that this expression is only sensible when all terms are non-negative and the top term is larger than the bottom one.

Next, we consider the relative phases of each term. Every time $k$ is incremented by one, there will be an additional factor of $i^2=-1$.  Putting these together, we can find the probability amplitude for outputting an equal number of photons
\begin{equation}
C_\mathrm{equal}(n,m)=\frac{i^{\frac{n+3m}{2}}\left(\frac{n+m}{2}\right)!}{\sqrt{n!m!2^{n+m}}}\sum_{k=\frac{n-m}{2}}^{\frac{n+m}{2}}\binom{n}{k}\binom{m}{k-\frac{n-m}{2}} (-1)^k. \label{eq:C_equal}
\end{equation}
We now use the symmetry of the binomial distribution, namely $\binom{n}{k}=\binom{n}{n-k}$. Using this symmetry, we can rewrite equation  \ref{eq:C_equal} as 
\begin{align}
C_\mathrm{equal}(n,m)=\frac{i^{\frac{n+3m}{2}}\left(\frac{n+m}{2}\right)!}{\sqrt{n!m!2^{n+m}}}\Bigg[\sum_{k=\frac{n-m}{2}}^{\lfloor \frac{n}{2} \rfloor}\binom{n}{k}\binom{m}{k-\frac{n-m}{2}} (-1)^k+ \nonumber \\
\sum_{k=\frac{n-m}{2}}^{\lfloor \frac{n}{2} \rfloor}\binom{n}{k}\binom{m}{k-\frac{n-m}{2}} (-1)^{m-k} +\delta_{\mathrm{mod}_2(m)=0}\binom{n}{\frac{n}{2}}\binom{m}{\frac{m}{2}} (-1)^{\frac{m}{2}}\Bigg].
\label{eq:C_equal_sym}
\end{align}
By writing the expression in this form, we see that if $n$ is odd (which also implies that $m$ must be odd for an equal sum to be possible), these will cancel termwise and therefore $C_\mathrm{equal}(n,m)=0$. If $n$ is even, such cancellation will not occur, both because $(-1)^{m-k}=(-1)^k$ and because there will be an overall odd number of combinatorial terms, with an unmatched one corresponding to $k=\frac{n}{2}$. Numerically, this can be seen by comparing figure \ref{fig:number_difference_odd} and \ref{fig:number_difference_even} .

An advantage of the photon number always being unequal in this case is that we can create states for which the encoded value using the DiPNE method (equation \ref{eq:bin_encode_photon}) always takes a well defined value,  since photon number in the two modes are strictly forbidden from being equal.

\subsection{Gaussian Driving\label{sub:Gauss_drive}}

\begin{figure}[h!]
    \centering
    \begin{subfigure}{0.48\linewidth}
        \centering
        \includegraphics[width=\linewidth]{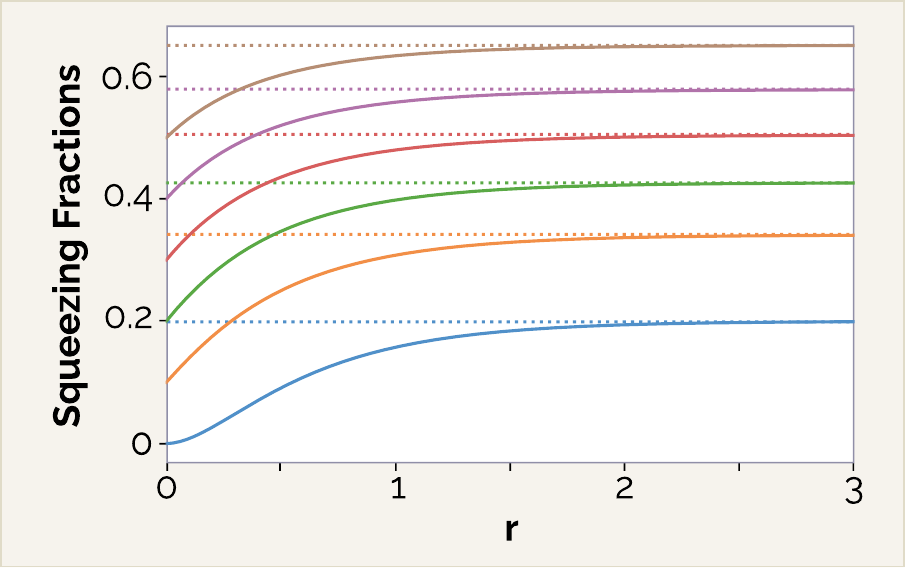}
        \caption{Change with antisqueezing}
        \label{fig:sq_frac_r}
    \end{subfigure}
    \hfill
    \begin{subfigure}{0.48\linewidth}
        \centering
        \includegraphics[width=\linewidth]{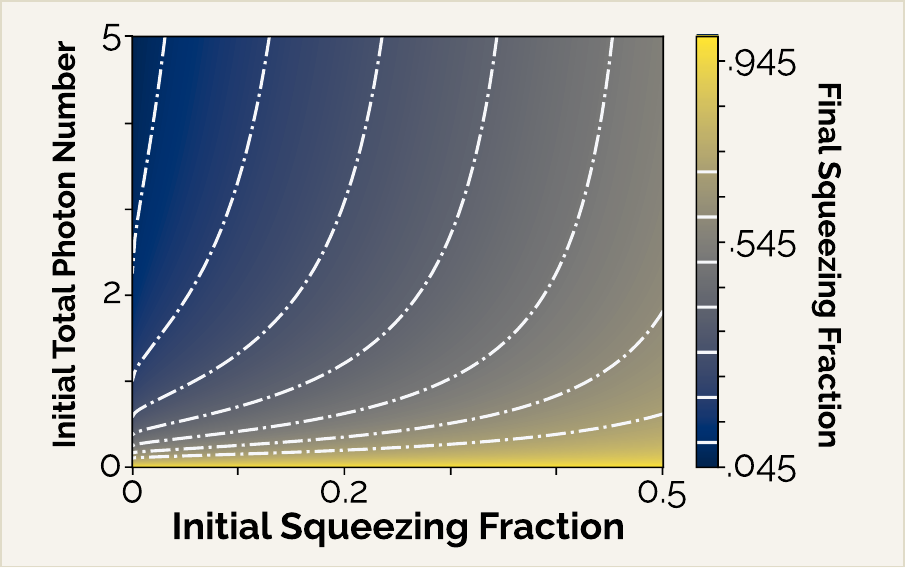}
        \caption{Strong antisqueezing limit}
        \label{fig:sq_frac_inf_r_0_n}
    \end{subfigure}
    \caption{Fraction for photons from squeezing effects after performing further antisqueezing on an antisuqeezed coherent state. a) after performing an antisqueezing of $r$ on an initial state with a total of one photon combined between antisqueezing and displacement. The dashed lines represent the strong squeezing limit. b) After strong antisqueezing is performed with a given initial number of photons and fraction which are from squeezing. These results are the created based on analytical calculations of the photon numbers and using the approximation in equation \ref{eq:frac_sq_strong approx} for the strong antisqueezing limit.}
    \label{fig:match_squeeze_excess}
\end{figure}

While in principle one could continue injecting increasingly high displacement superpositions by balancing squeezing and beamsplitting angle, it may be easier in practice to drive in a Gaussian way, once a non-Gaussian superposition has been established.  Some options are as follow:

\begin{enumerate}
\item Directly squeeze the modes used for computation: \\
\\
Applying squeezing to a mode which is already displaced will amplify the displacement. This is very similar, but not identical, to how a coherent Ising machine operates, but with the two coherent states in separate modes, rather than opposite quadratures on the same mode.  It is worth noting that even at very large squeezing values, the fraction of photons from displacement will approach a finite (non-zero) value. If we consider a state which starts with (real) displacement $d_0$ and anti-squeezing $r_0$, then the fraction of photons from squeezing effects after a large anti-squeezing of $r$ will be \cite{Gerry_Knight_2004,brask2022gaussianstatesoperations}:
\begin{equation}
\frac{d^2_0\exp(2r)}{\sinh^2(r+r_0)}\approx \frac{4 d^2_0}{\exp(2r_0)}. \label{eq:frac_sq_strong approx}
\end{equation}
Even a very conservative estimate with single photon subtraction (compare with rough values in \ref{fig:kitten_squeeze}) of a $d_0=1$ (a single photon of initial displacement) and $0.1$ photons of anti-squeezing ($r_0=\sinh^{-1}(\sqrt{0.1})$), the number of photons from squeezing would be less than half the number from displacement (or a fraction of photons from squeezing of less than $\frac{1}{3}$ if reported in the way shown in  figure \ref{fig:kitten_squeeze}). We show the convergence to the strong-squeezing limit for different initial conditions in figure \ref{fig:sq_frac_r}, and a heatmap of the final fraction of photons from squeezing effects in this limit in figure \ref{fig:sq_frac_inf_r_0_n}.
\item Injections of coherent states into each mode:\\
\\
Another way to drive is to inject coherent states into each mode using the scheme depicted in figure \ref{fig:kit_inject} but using two unentangled coherent states as input.  This has the advantage of being able to reinforce the phase relationships between the modes. Interference effects also means that more photons will be transferred into modes which already have a large number of photons. There is no analog to this kind of driving in the coherent Ising encoding.
\item Squeezing kitten states before injection: \\
\\
Because anti-squeezing performed on a coherent state increases displacement, squeezing a kitten state before injection will increase the displacement of each mode in the (approximate) superposition. This can be particularly useful to match the displacements when different $k$ values are measured in \ref{fig:kit_create}. While mixing odd and even kitten states is likely to lead to complications in terms of relative phase, one could still do this with every odd kitten state.  If it is possible to measure an arbitrary number of photons, this could lead to a useful state for injection being produced roughly half of the time.
\end{enumerate}

\begin{figure}[h!]
    \centering
    \begin{subfigure}{0.48\linewidth}
        \centering
        \includegraphics[width=\linewidth]{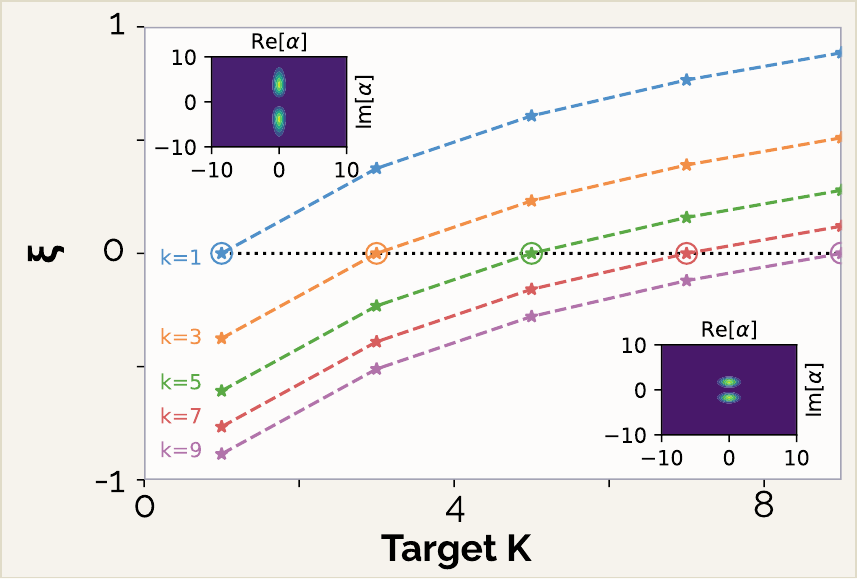}
        \caption{Anti-squeezing Required}
        \label{fig:match_squeezing}
    \end{subfigure}
    \hfill
    \begin{subfigure}{0.48\linewidth}
        \centering
        \includegraphics[width=\linewidth]{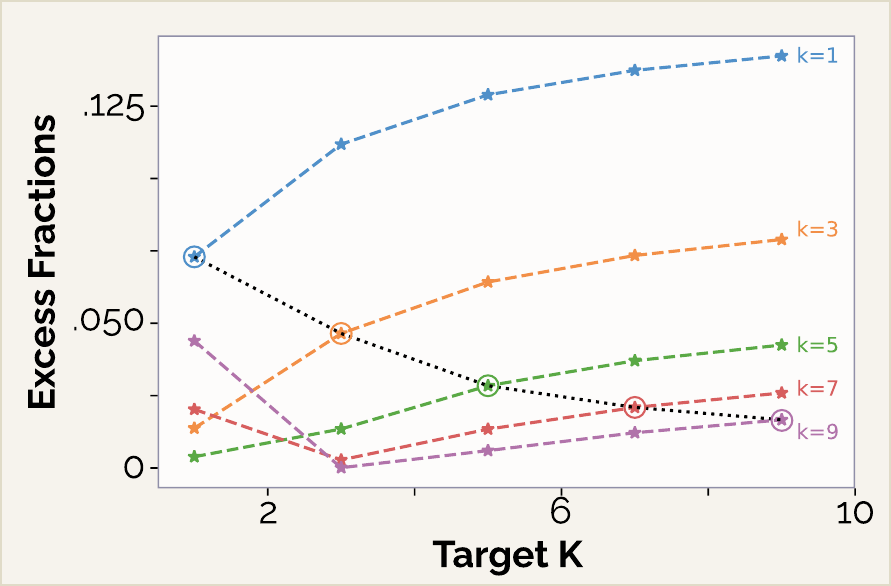}
        \caption{Fractional Excess Photons}
        \label{fig:excess_photons_match}
    \end{subfigure}
    \caption{Quantities related to matching displacements with kitten states produced by measuring an initial $k$ photons in the infinite squeezing limit.  The initial $k$ values are labeled on the plots.  Cases where the initial $k$ is already the target and therefore no squeezing is required are circled and joined with a dotted curve which acts as a guide to the eye.  Subfigure (a) shows the amount of squeezing required to match the displacement.  Subfigure (b) shows the fraction of photons which are not required in a purely displaced state (one minus the fraction required for displacement divided by the actual average photon number). The insets of (a) show the Husimi-Q function after (anti-)squeezing for the two extreme cases.  The bottom plot shows the Q distribution for the state produced with $9$ photons measured squeezed to match the state with one,  and the upper the result of squeezing the state resulting from $1$ photon detection squeezed to match with the result of $9$ photons being measured. Calculations performed in a truncated Hilbert space with up to $1000$ photons.}
    \label{fig:match_squeeze_excess}
\end{figure}

Note that while methods 1 and 3 will generally induce (anti-)squeezing to the state, we have already argued that this will not have an effect on the photon loss rates due to interference and therefore will not disrupt a valid encoding.  In figure \ref{fig:match_squeeze_excess} we see some results for this process.  Figure \ref{fig:match_squeezing} shows the amount of squeezing required to match displacements,  for all odd kitten states resulting from $1$ to $9$ photons measured.  Figure \ref{fig:excess_photons_match} shows the excess amount of photons beyond what would be required for a cat state with the same displacement,  note that because the kitten states we produce have some squeezing in addition to a small amount of non-Gaussian fluctuations in the superimposed mode, this is not zero even when no squeezing is performed,  and in some cases matching a different $k$ value ends up decreasing the excess fraction since it would remove squeezing photons which are already present.  In any case the highest observed value is only about $14\%$ of the photons,  and less than $10\%$ in all cases other than starting with $k=1$.

\section{Numerical methods\label{sec:num_meth}}

All numerical calculations were performed using Python.  The numpy \cite{numpy} and scipy \cite{2020SciPy-NMeth} package was heavily used for calculations and matplotlib \cite{hunter2007matplotlib} for visualization. Jupyter notebooks \cite{jupyter} were also used. All studies reported here used a truncated full state space simulation, with variable truncation levels given in the captions of respective figures.  Sometimes squeezed states were initialized using an analytical formula rather than numerical matrix exponentiation. This technique reduced numerical artifacts related to the photon number cutoff and is stated explicitly where used.

\section{Discussion and Conclusions\label{sec:disc_conc}}

In this manuscript, we have presented analysis on if and how truly quantum analog optimiser, which uses direct encoding, on similar footing to a quantum annealer, could be engineered. The conclusion is that the fundamental components for such an optimiser does indeed exist. In particular, displacements provide a convenient subspace which allows for parallel computation. We find that the loss which encodes the problem is insensitive to squeezing as well as a large class of non-Gaussian fluctuations.  We observe that unlike previous literature which focuses only on preparing superpositions of coherent states, if we instead allow some squeezing in the superposition, then standard Schr{\"o}dinger kitten state preparation methods can prepare the necessary superpositions at very high fidelity.  We usually exceed $99\%$ fidelity with these states and even exceed $99.9\%$ in many cases.  Even in these cases, squeezing often accounts for less than $10\%$ of the photons, indicating that squeezing is unlikely to overwhelm the underlying signal from displacement. This also admits the possibility of the displacements in superposition states used in driving to be standardized via squeezing after preparation. This would allow many more of the produced kitten states to be used in practice, rather than waiting for specific measurement results.

We further find that, at least at a theoretical level, it is possible to avoid decoherence from the light that is lost due to interference. Furthermore, degrees of freedom which need to be standardized can be measured without disrupting the core information being processed. 

The aim of this work is a highly theoretical one, to establish the foundations of whether a fully-quantum all-optical analog optimiser is possible and enumerating what the key considerations are. We have proposed closely related DiPNE and DiDE encoding strategies, based on the relative photon numbers and displacements in two modes, respectively. This is important to lay the foundations for developing new protocols. 

\appendix

\section*{Appendices}

\section{Interference for symmetric displaced states \label{app:int_gen}}

Let us consider a general two-mode quantum state $\ket{\zeta}$ which has a mean displacement of zero in the quadratures of both variables; in other words,
\begin{equation}
\sandwich{\zeta}{\hat{a}_0+\hat{a}^\dagger_0}{\zeta}=\sandwich{\zeta}{\hat{a}_1+\hat{a}^\dagger_1}{\zeta}=\sandwich{\zeta}{\hat{a}_0-i\hat{a}^\dagger_0}{\zeta}=\sandwich{\zeta}{\hat{a}_1-i\hat{a}^\dagger_1}{\zeta}=0.
\end{equation}
This equation further implies that the displacement in any direction will be zero,
\begin{equation}
\sandwich{\zeta}{x_0\hat{a}_0+x^\star_0\hat{a}^\dagger_0+x_1\hat{a}_1+x^\star_1\hat{a}^\dagger_1}{\zeta}=0, \label{eq:2mode_no_disp}
\end{equation}
for any arbitrary complex values of $x_0$ and $x_1$.

We define an arbitrary displacement applied to this state,
\begin{align}
\hat{D}(\alpha_0,\alpha_1)=\exp\left( \alpha _0\hat{a}^\dagger_0-\alpha_0^\star\hat{a}_0 \right)\exp\left( \alpha _1\hat{a}^\dagger_1-\alpha_1^\star\hat{a}_1 \right)= \nonumber \\
\exp\left(-\frac{|\alpha_0|+|\alpha_1|}{2} \right)\exp\left( \alpha _0\hat{a}^\dagger_0\right)\exp\left(- \alpha^\star _0\hat{a}_0\right)\exp\left( \alpha _1\hat{a}^\dagger_1\right)\exp\left(- \alpha^\star _1\hat{a}_1\right),
\end{align}
where the second line follows from the disentangling theorem \cite{Gerry_Knight_2004} and the fact that operators commute when acting on different Hilbert spaces.  We now consider displacements to our arbitrary undisplaced state
\begin{equation}
\ket{\zeta_d}=\hat{D}(\alpha_0,\alpha_1)\ket{\zeta}.
\end{equation}
We are interested in the final number of photons found in mode $1$ after a beamsplitting operation
\begin{equation}
n_f=\sandwich{\zeta_d}{\hat{B}^\dagger(\theta)\hat{a}^\dagger_1\hat{a}_1\hat{B}(\theta)}{\zeta_d},
\end{equation}
where $\hat{B}(\theta)$ is an arbitrary beamsplitter parameterized by angle $\theta$.  By applying the definition of the beamsplitting operation and commuting it through the displacement operator, we have 
\begin{align}
\hat{B}(\theta)\ket{\zeta_d}=\hat{B}(\theta)\hat{D}(\alpha_0,\alpha_1)\ket{\zeta}= \nonumber \\
\hat{D}(\cos(\theta)\alpha_0+i\sin(\theta)\alpha_1,\cos(\theta)\alpha_1+i\sin(\theta)\alpha_0)\hat{B}(\theta)\ket{\zeta}. \label{eq:beamsplit_commute}
\end{align}
We then consider the combination of a displacement and annihilation operator, (again using the disentangling theorem \cite{Gerry_Knight_2004})
\begin{align}
\hat{a}\hat{D}(\alpha)=\hat{a}\exp\left( \alpha \hat{a}^\dagger-\alpha^\star \hat{a} \right)= \hat{a}\exp\left( \alpha \hat{a}^\dagger \right)\exp\left(-\frac{|\alpha|}{2} \right) \exp\left(-\alpha^\star \hat{a} \right)= \nonumber \\
\hat{a}\sum^\infty_{n=0}\frac{\left( \alpha \hat{a}^\dagger \right)^n}{n!}\exp\left(-\frac{|\alpha|}{2} \right) \exp\left(-\alpha^\star \hat{a} \right), \label{eq:a_disp_expand}
\end{align}
where we have performed a series expansion in $\hat{a}^\dagger$ on the last line since those are the terms which do not commute with $\hat{a}$. We now make use of the commutation relationship $[\hat{a},\hat{a}^\dagger]=1$, which further implies that 
\begin{equation}
[\hat{a},\left(\hat{a}^\dagger\right)^n]=n\left(\hat{a}^\dagger\right)^{n-1}.
\end{equation}
Applying this commutator termwise to the series expansion yields 
\begin{align}
\hat{a}\sum^\infty_{n=0}\frac{\left( \alpha \hat{a}^\dagger \right)^n}{n!}=\sum^\infty_{n=0}\frac{\left( \alpha \hat{a}^\dagger \right)^n}{n!}\hat{a}+\sum^\infty_{n=0}\frac{n\alpha^n \left( \hat{a}^\dagger \right)^{n-1}}{n!}= \nonumber \\
\sum^\infty_{n=0}\frac{\left( \alpha \hat{a}^\dagger \right)^n}{n!}\hat{a}+\sum^\infty_{n=0}\frac{\alpha^n \left( \hat{a}^\dagger \right)^{n-1}}{(n-1)!}=\sum^\infty_{n=0}\frac{\left( \alpha \hat{a}^\dagger \right)^n}{n!}\left(\hat{a}+\alpha\right).
\end{align}
Plugging this result into equation \ref{eq:a_disp_expand} yields
\begin{align}
\hat{a}\hat{D}(\alpha)=\hat{D}(\alpha)\hat{a}+\alpha\hat{D}(\alpha).
\end{align}
We can then further use equation \ref{eq:beamsplit_commute} to show that 
\begin{align}
\hat{a}_1\hat{B}(\theta)\ket{\zeta_d}=
\hat{D}(\cos(\theta)\alpha_0+i\sin(\theta)\alpha_1,\cos(\theta)\alpha_1+i\sin(\theta)\alpha_0)\nonumber \\ \left(\hat{a}_1+\cos(\theta)\alpha_1+i\sin(\theta)\alpha_0\right)\hat{B}(\theta)\ket{\zeta}.
\end{align}
Using the fact that displacements are unitary and combining this equation with it's Hermitian conjugate we have
\begin{align}
n_f=\nonumber \\
\left|\cos(\theta)\alpha_1+i\sin(\theta)\alpha_0\right|^2\sandwich{\zeta}{\hat{B}^\dagger(\theta)\hat{B}(\theta)}{\zeta}+\sandwich{\zeta}{\hat{B}^\dagger(\theta)\hat{a}^\dagger_1\hat{a}_1\hat{B}(\theta)}{\zeta}+ \nonumber \\
\sandwich{\zeta}{\hat{B}^\dagger(\theta)\left(\cos(\theta)\alpha^\star_1\hat{a}_1+\cos(\theta)\alpha_1\hat{a}^\dagger_1-i\sin(\theta)\alpha^\star_0\hat{a}_1+i\sin(\theta)\alpha_0\hat{a}^\dagger_1\right)\hat{B}(\theta)}{\zeta}.
\end{align}
The second line of this expression is identically zero due to equation \ref{eq:2mode_no_disp}, and furthermore by normalisation of $\ket{\zeta}$ and the unitarity of beamsplitting the first expectation is equal to one.  The average photon number in mode $1$ at the end of the process therefore simplifies to 
\begin{align}
n_f=
\left|\cos(\theta)\alpha_1+i\sin(\theta)\alpha_0\right|^2+\sandwich{\zeta}{\hat{B}^\dagger(\theta)\hat{a}^\dagger_1\hat{a}_1\hat{B}(\theta)}{\zeta},
\end{align}
which consists of a sum of a number of displacement photons, which does not depend on the structure of $\ket{\zeta}$ and a component which depends on this structure but not any of the displacement.  We have therefore shown that the displacements will act separately from the internal structure of the displaced state.  Based on this fact we should expect the loss due to interference will be well controlled even in the presence of additional non-Gaussian modifications on the displaced states.

An astute reader may wonder if relative phases in interference could effect the value of $\sandwich{\zeta}{\hat{B}^\dagger(\theta)\hat{a}^\dagger_1\hat{a}_1\hat{B}(\theta)}{\zeta}$. While a completely general argument which holds for all $\ket{\zeta}$ is not possible here, we can make an argument for a large class of states. Specifically, if 
\begin{align}
\exp\left(i \pi\hat{a}^\dagger_1\hat{a}_1\right)\ket{\zeta}= \pm\ket{\zeta}\,\, \mathrm{or}  \\
\exp\left(i \pi \hat{a}^\dagger_0\hat{a}_0\right)\ket{\zeta}= \pm\ket{\zeta} \label{eq:sym_cond}
\end{align}
These states include those where only an odd or even number of photons are allowed in either of the two modes, or equivalently those with odd or even symmetry under a $\pi$ phase rotation.   

Since coherent states form an (over-)complete basis for optical states, any quantum state can be written as
\begin{equation}
\ket{\zeta}=\int d^2\alpha_0 \int d^2\alpha_1 C(\alpha_0,\alpha_1) \ket{\alpha_0}\ket{\alpha_1}
\end{equation}
where the square in the differential is to remind us that the integration is over the complex plane. Since $\exp\left(i \pi \hat{a}^\dagger_0\hat{a}_0\right)\ket{\alpha}=\ket{-\alpha}$, the condition in equation \ref{eq:sym_cond} puts a constraint on the values of $C$, namely for it to hold it must be true that
\begin{equation}
C(\alpha_\zeta,\alpha_1)=\pm C(-\alpha_\zeta,\alpha_1)
\end{equation}
A further generalisation is to consider additional displacement on mode $0$, in which case,  based on the linearity of displacements, if we displace mode $0$ by $\alpha_0$ we have
\begin{equation}
C(\alpha_0+\alpha_\zeta,\alpha_1)=\pm C(\alpha_0-\alpha_\zeta,\alpha_1)
\end{equation}

Now let us consider a pair of the coherent states in this definition with opposite phase
\begin{align}
\ket{\psi_\mathrm{pair}}=C(\alpha_0+\alpha_\zeta,\alpha_1)\ket{\alpha_0+\alpha_\zeta}\ket{\alpha_1}+C(\alpha_0-\alpha_\zeta,\alpha_1)\ket{\alpha_0-\alpha_\zeta}\ket{\alpha_1}=\nonumber \\
C(\alpha_0+\alpha_\zeta,\alpha_1)\left(\ket{\alpha_0+\alpha_\zeta}\ket{\alpha_1} \pm \ket{\alpha_0-\alpha_\zeta}\ket{\alpha_1} \right) \rightarrow  \nonumber \\ C(\alpha_0+\alpha_\zeta,\alpha_1) \nonumber \\ \big(\ket{\cos(\theta)(\alpha_0+\alpha_\zeta)+i\sin(\theta)\alpha_1}\ket{\cos(\theta)\alpha_1+i\sin(\theta)(\alpha_0+\alpha_\zeta)} \pm \nonumber \\ 
\ket{\cos(\theta)(\alpha_0-\alpha_\zeta)+i\sin(\theta)\alpha_1}\ket{\cos(\theta)\alpha_1+i\sin(\theta)(\alpha_0-\alpha_\zeta)}\big).
\end{align}
When calculating photon numbers, the cross terms involving $\alpha_\zeta$ and other displacements will cancel, explicitly we have (assuming the output modes are labelled $2$ and $3$)
\begin{align}
\frac{\sandwich{\psi_\mathrm{pair}}{\hat{a}^\dagger_2\hat{a}_2}{\psi_\mathrm{pair}}}{|C(\alpha_0+\alpha_\zeta,\alpha_1)|^2}=|\cos(\theta)\alpha_0+i\sin(\theta)\alpha_1|^2+|\cos(\theta)\alpha_\zeta|^2 \mathrm{and} \\
\frac{\sandwich{\psi_\mathrm{pair}}{\hat{a}^\dagger_3\hat{a}_3}{\psi_\mathrm{pair}}}{|C(\alpha_0+\alpha_\zeta,\alpha_1)|^2}= |\cos(\theta)\alpha_1+i\sin(\theta)\alpha_0|^2+|\sin(\theta)\alpha_\zeta|^2.
\end{align}
where we observe that $\alpha_\zeta$ only appears by itself in modulus squared terms and therefore will contribute photons to each output mode in a way which is independent of phase relationships with other modes and its contribution to the photon number in each mode depends only on the beamsplitting angle. Since this argument holds individually for each pair,  it must also hold overall. Note in particular that we have assumed nothing about the state in mode $1$, so the argument would hold if it were of a similar form. While not strictly convering all displacement-free states\cite{footnote3} equation  \ref{eq:sym_cond} contains a number which are frequently observed in practice, including squeezed vacuuum,  and non-Gaussian states including both cat-like and kitten states with arbitrary squeezing. 

\section{Background\label{app:background}}

There are at least two established paradigms for optical optimisation, coherent Ising machines and variational Gaussian Boson Sampling, in addition to the emerging entropy computing paradigm. To give context to the results in the main text, it is worth briefly reviewing the existing paradigms and discuss their strengths and weaknesses. Before reviewing these paradigms, we will briefly review a core concept, that of coherent, Gaussian and non-Gaussian states of light.

\subsection{Gaussian and coherent states of light}

A concept within quantum optics, which will be important for understanding the results and discussion presented later in this manuscript is that of Gaussianity and non-Gaussianity. These concepts play an important role in how easy a system is to classically simulate.  A system which is easy to simulate classically cannot deliver a true quantum advantage, as the simulation could simply be used in place of a physical system\cite{footnote4}. We start our discussion with coherent states, the quantum analog of classical electromagnetic waves. While not completely classical \cite{Gerry_Knight_2004}, coherent states act as electromagnetic waves with some uncertainty in the field and energy values. The mean displacement and phase of coherent states however transform as electromagnetic waves would be expected to under beamsplitting; in particular, performing beamsplitting on a pair of coherent states will not produce entanglement. From this observation alone, it is clear that by themselves coherent states alone cannot effectively be used for quantum information processing. This applies even to very weak coherent states, which may on average have a single photon or less \cite{Lee2019approxSinglePhoton}.

Field uncertainty can be reduced by squeezing operations, which periodically reduces the uncertainty in the fields at a particular point in time at the cost of increasing it at another \cite{Gerry_Knight_2004}. States produced by a combination of displacement (the operations which create coherent states) and squeezing produce a class of optical states known as Gaussian states  \cite{Bartlett2002GaussEfficient,Wang2007GaussQuantInfo}. Gaussian states can support quantum correlations; beamsplitting of separable Gaussian states can produce entanglement.  Gaussian states, however, are efficiently described by only the first and second moments of the quadrature operators, and therefore these degrees of freedom can be simulated and described in a classically efficient way. A procedural guide for how to do this can be found in \cite{brask2022gaussianstatesoperations}.  The photon counting statistics of Gaussian states of light cannot however be efficiently simulated, and non-Gaussian measurements, such as photon counting measurements, can render a Gaussian state non-Gaussian, and therefore no longer easily simulable, a process which will be important to later discussions.  Non-Gaussianity can also be induced by evolving with any operator generated from terms which involves a higher order than second order of any combination of creation/annihilation operator. As a concrete example, terms of the form $\hat{a}^{(\dagger)}_1\hat{a}^{(\dagger)}_2$ maintain Gaussianity, while those of $\hat{a}^{(\dagger)}_1\hat{a}^{(\dagger)}_2\hat{a}^{(\dagger)}_3$ would not, where $1$, $2$, and $3$ are arbitrary (possibly repeated) indices, and a dagger in parentheses indicates that an operator can be a creation or annihilation operator. For a review of non-Gaussianity and how it relates to other properies such as Wigner negativity, see \cite{Walschaers2021nonGauss}.

As a point of terminology, squeezing operations which are performed in a quadrature which is not the one which is measured can both increase uncertainty and amplify existing coherent states in the relevant quadrature. This type of squeezing is often referred to as anti-squeezing in the literature, we adopt this terminology as well. As we discuss later, anti-squeezing plays a fundamental role in the operation of a coherent Ising machine.

While it may seem counterintuitive that highly quantum dynamics can be efficiently simulated classically, an analogous set of operations exist in gate model quantum computing, such as the Clifford states which can also support a high degree of entanglement, but can be efficiently simulated using stabilizers, as long as no non-Clifford operations are performed \cite{gottesman1998GK}. Where existing methods sit and what is classically simulable is summarized in table \ref{tab:simulability}.

\begin{table}
\begin{centering}
\begin{tabular}{|c|c|c|c|}
\hline
 & Measurements $\Rightarrow$ & Gaussian & non-Gaussian \tabularnewline
\hline
\hline
\multirow{3}{*}{\vspace{-0.2cm}\rotatebox[origin=c]{90}{\parbox[c]{1.5 cm}{\centering Input States and Operations}}}& Coherent states+Beamsplitters & simulable \cellcolor{red!25} & simulable\cellcolor{red!25} \tabularnewline
\cline{2-4}
 & Gaussian states+Beamsplitters & simulable \cellcolor{red!25} & Gaussian Boson Sampling \cellcolor{green!25} \tabularnewline
\cline{2-4}
 & non-Gaussian & Coherent Ising\cellcolor{green!25} &  Boson Sampling \cellcolor{green!25} \tabularnewline
\hline
\end{tabular}
\par
\end{centering}
\caption{Table of established methods in terms of the states/dynamics and measurement types. Colour coding is used to indicate classical simulability of the full system, with red indicating classical simulability and green indicating a lack of simulability. It is worth noting that this table assumes an all optical coherent Ising machine, as frequent measurements would likely render the measurement and feedback version efficiently classically simulable. The final row is equivalent for dynamics under non-Gaussian Hamiltonians and/or non-Gaussian initial states.  This table also refers to simulating the full dynamics, it is possible that a Gaussian simulation of a coherent Ising machine could capture the relevant behaviours for optimisation without being able to fully simulate the system. Note also, this is referring to an all-optical coherent Ising machine, as the measurement-and-feedback model will be classically simulable. Note also that this assumes that loss of coherence due to lack of quantum erasure does not render the device easily simulable, a discussion which is far beyond the scope of this paper. \label{tab:simulability}}
\end{table}

\subsection{Gaussian Boson Sampling as a variational optimiser}

Boson sampling is a quantum paradigm which leverages the complexity of photon counting statistics.  The advantage of this paradigm is that it can sample distributions which are strongly suspected to be hard to sample from by any classical algorithm.  The original proposal for Boson sampling was to input Fock states into a mesh of beamsplitters and measure the photon numbers at the output \cite{scheel2004permanents,Aaronson2013BosonSampling}. Although the results for performing such experiments with classical electromagnetic waves, or coherent states, which are their closest quantum optical analog are easy to predict \cite{Gerry_Knight_2004}, the results are no longer classically simple when Fock states are used as input.  Effectively, Boson sampling leverages the Bosonic statistics of photons, typified in simple cases by the Hong-Ou-Mandel effect \cite{Hong1987HOM}, where a pair of identical photons are observed to bunch together when incident on a beamsplitter, to sample from complex distributions. While the mathematical details are not important here, the sampled distribution corresponds to the permanent of a matrix \cite{scheel2004permanents,Aaronson2013BosonSampling}, an analog of the determinant which does not have the alternating sign structure. Concretely the origin of the permanent comes from the fact that (following the derivation in \cite{scheel2004permanents}), a quantum state being acted on by a unitary beamsplitting operator $\hat{U}$ represented by a matrix $\Lambda$ can be written as 
\begin{equation}
\hat{U}\ket{n_1,n_2....n_N}=\prod_{i=1}^N\frac{1}{\sqrt{n_i!}}\left(\sum_{k_i=1}^N\Lambda_{k_i,i}\hat{a}^\dagger_{k_i} \right)^{n_i}\ket{0}^{\otimes N}
\end{equation}
taking an inner product with $\bra{m_1,m_2...m_N}=\bra{0}^{\otimes N}\prod_{i=1}^N(\hat{a}_i)^{m_i}$ then yields
\begin{align}
\bra{m_1,m_2....m_N}\hat{U}\ket{n_1,n_2....n_N}=\nonumber \\
\prod_{j=1}^N\prod_{i=1}^N\frac{1}{\sqrt{n_i!}}\frac{1}{\sqrt{m_j!}}\bra{0}^{\otimes N}\hat{a}_j^{m_j}\left(\sum_{k_i}^N\Lambda_{k_i,i}\hat{a}^\dagger_{k_i} \right)^{n_i}\ket{0}^{\otimes N}= \\
\prod_{j=1}^N\prod_{i=1}^N\frac{1}{\sqrt{n_i!}}\frac{1}{\sqrt{m_j!}}\sum_{\sigma \in S_{ N_{p}}}\prod_{i=1}^{N_p}\Lambda^{(\mathrm{sub)}}_{i,\sigma_i}=\prod_{j=1}^N\prod_{i=1}^{N}\frac{1}{\sqrt{n_i!}}\frac{1}{\sqrt{m_j!}}\mathrm{Per}(\Lambda^{(\mathrm{sub)}})
\end{align}
where $N_p$ is the total number of photons, $S$ is the permutation group, $\Lambda^{\mathrm{(sub)}}$ is a matrix of occupied modes\cite{footnote5} (with repeated rows for multiple photon occupation), and $\mathrm{Per}$ is the permanent.

Since the determinant is invariant under basis rotations, it can be efficiently calculated by diagonalising and taking the product of the eigenvalues. The permanent however is not basis invariant, so it has to be calculated in the original basis; the naive formula requires a super-exponentially (factorially) growing number of terms, and finding the exact value of the permanent is known to be \#P hard \cite{Valiant1979PermanentHard}. Even approximating the permanent is suspected to be hard \cite{Cai1999PermanentHard,Aaronson2013BosonSampling}.

While the original Boson sampling formulation is (relatively) simple and elegant, reliably preparing many Fock states with indistinguishable photons is a challenging task. An experimentally simpler task is to prepare squeezed quantum vacuum states, with measurements still performed in the Fock basis. This paradigm is known as Gaussian Boson sampling. Gaussian Boson sampling does not sample from distributions defined by permanents, but rather a related quantity known as a Hafnian \cite{footnote6,Hamilton2017GBS,Kruse2019GBS}.  The computation of a permanent can be reduced to an analogous Hafnian calculation of a larger matrix; therefore, the sampling from a distribution defined in terms of Hafnians inherits the hardness of sampling from distributions defined by permanents. 

While strongly suspected to have a formal quantum advantage (in terms of classical difficulty of the underlying problem), Gaussian Boson sampling does not directly map to optimisation in any known way. Gaussian Boson sampling can still be used as a tool for optimisation by using it within variational techniques \cite{bradler2021ORCA,goldsmith2024ORCA_TSP}. Conceptually, variational quantum algorithms use a quantum device to sample from a rich and expressive statistical distribution and classically compute the quality of each sample.  Based on the relative quality of the samples, different parameter updates can be performed to iteratively improve the results. A disadvantage of this variational approach is that the problem of interest is never mapped directly to the device. 

Quantum variational algorithms are known to have a tradeoff between expressivity and trainability.  In other words, if the device parameters allow too much control over the final distribution,  then small changes will not change the quality of the results meaningfully, a phenomena known as ``barren plateaus'' \cite{McArdle2020variational,Holmes2022Barren,Larocca2022diagnosingbarren,Callison2022hybrid}. On the other hand, a distribution that is not expressive enough may entirely fail to find high-quality solutions. Variational algorithms are still promising and an area of active research, but this tradeoff presents a unique disadvantage which is not present in other techniques. In the implementation of variational Gaussian Boson Sampling we are aware of, binary values are usually mapped to the parity (odd or evenness) of photon number in each output \cite{bradler2021ORCA,goldsmith2024ORCA_TSP} and the variation is performed over beamsplitting angle \cite{goldsmith2024ORCA_TSP}.

\subsection{Coherent Ising machine}

An alternative technique for performing optimisation optically is the coherent Ising machine \cite{Inagaki16a,Yamamoto2017CoherentIsing,Yamamoto2020CoherentIsing,Honjo2021CoherentIsing,Mosheni2022IsingMachines,Kumagai2025singlePhotonCIM}.   As the name suggests, these devices are able to directly encode an Ising optimisation problem. However, unlike (Gaussian) Boson sampling, the computational complexity theory aspects related to the role of quantum mechanics are less clear. The largest and most performant Coherent Ising machines use classical measurement and feedback in a way which prevent large-scale quantum superpositions \cite{Yamamoto2017CoherentIsing}. While an interesting area in general, our present work is concerned with using quantum effects in a way which cannot be easily simulated, so we will focus on all-optical proposals, which do not involve such feedback. Within the main text of this work, it is implicit that whenever we refer to a coherent Ising machine we mean an \emph{all optical} coherent Ising machine \cite{Yamamoto2017CoherentIsing,Stranati2021opticalCIM,Bo2025OpticalCIM}.   

We now briefly review how they work. The key concept is that anti-squeezing can drive an optical mode in a way which amplifies existing displacements along the anti-squeezing quadrature axis with no inherent preference toward displacement in either a positive or negative direction \cite{Yamamoto2020CoherentIsing}. A classical analogy to this kind of driving is a swing where a well timed increase or decrease of the moment of inertia can amplify existing small oscillations in one of two phases, but will not start oscillation on its own \cite{Yamamoto2020CoherentIsing}. Interference, however, is phase dependent; destructive interference on the channel that is not sent to erasure leads to higher loss. The coherent Ising machine thus iteratively searches a solution space by repeatedly performing anti-squeezing and interference.  Based on general physical arguments, there will be a point where the loss and gain (via anti-squeezing) match each other, and this happens first for the least-lossy (most optimal) state.  When only the least-lossy state has gain that can exceed the loss, then the device should search the space until that optimal configuration is found. For this state, the gain exceeds the loss, and if found, the system will not leave it as progressive amplification will make it very difficult to move into a different configuration. In practice, the value of gain where the least-lossy state starts to be amplified is not known, so instead, the anti-squeezing strength is swept slowly, in the hope that the optimal solution can be found before less optimal states are amplified.

At face value, the model discussed in the previous paragraph does perform some kind of parallel search, and the anti-squeezing creates a kind of superposition of displacement in positive and negative directions for each Ising variable. However, in this simple model, this cannot be a true quantum parallel search. The process is completely Gaussian and can be efficiently classically simulated; such simulations are interesting computational tools in their own right and are known as simulated bifurcation machines\cite{Clements2017GaussIsing,Tiunov2019simIsing,Tatsumura2021simBif,Ng2022GaussIsing}.  Recall that the simulability will remain even in the few-photon regime studied in \cite{Kumagai2025singlePhotonCIM}.

If we consider a more detailed model of the coherent Ising machine, there are sources of non-Gaussianity. The two that have been identified are photon loss and saturation of the non-linear crystal that is used for squeezing \cite{Yamamoto2017CoherentIsing}. We will focus on the latter in this discussion, as we are not aware of any work which has built on the photon loss argument. In practice, anti-squeezing is performed by using a degenerate optical parametric oscillator. Effectively, this is implemented by a non-linear material $\chi^{(2)}$, where two of the modes are degenerate. The Hamiltonian in this case takes the form 
\begin{equation}
\hat{H}_{NL}=i \chi^{(2)}\left(\hat{a}_p\hat{a}^\dagger\hat{a}^\dagger-\hat{a}^\dagger_p\hat{a}\hat{a}\right)
\end{equation}
where $\hat{a}_p$ is the pump mode annihilation operator and $\chi^{(2)}$ is a second-order non-linear susceptibility (we work in natural units where $\hbar$ has been absorbed). This Hamiltonian involves third-order terms, and so is non-Gaussian.  The usual assumption to construct a squeezing operator is to assume that the pump mode is very strong in comparison to the coupling; hence, the pump mode is approximately unchanged by the non-linear interaction and the field operators can then be replaced with a complex constant $\chi^{(2)}\hat{a}_p\rightarrow \eta$, in which we can recover the (Gaussian) squeezing Hamiltonian\cite{Gerry_Knight_2004}
\begin{equation}
\hat{H}_\mathrm{squeeze}=i\eta \hat{a}^\dagger\hat{a}^\dagger-i\eta^\star\hat{a}\hat{a}.
\end{equation}
If a situation is engineered where this assumption is not valid, the system becomes non-Gaussian, but the question then becomes whether this non-Gaussianity plays an important role in the computation; hence, we ask whether it presents any advantage over the simulated bifurcation picture. In \cite{Yamamura2017quantumIsing}, the authors showed that such saturation may produce some features reminiscent of quantum superposition, including interference fringes in Wigner functions, but it is unclear the extent to which these allow quantum parallelism to be exploited. 

Another issue, which even persists in the all optical setting and for which we are not aware of any discussion within the literature is the fact that the lost photons in all versions of coherent Ising machines we are aware of are removed in an uncontrolled way, presumably either to a beam dump or the substrate of a chip in practice. This is potentially problematic because this light carries information about the state of the system. In particular, by energy conservation, more light will leave the system in the case of destructive interference on the mode which is fed back.  In fact, even the total amount of photon loss, which can be expressed macroscopically in the change of temperature of the area where the energy is deposited carries information on the optimality of a given solution candidate. This will disrupt superpositions between more and less optimal states; hence, even this level of information leads to decoherence. An analysis of the severity of this problem is beyond the scope of this manuscript, but it is worth highlighting. It is also worth highlighting that this decoherence is not a fundamental issue, at least from a theoretical perspective. There will be methods to erase this information, which will effectively hide such variations under the natural variance of a strong coherent beam of light; we discuss this in more detail in section  \ref{sub:quant_erase}. We discuss this in context of a displacement-based encoding, but it will naturally extend to the phase encoding used in the coherent Ising setting.

\subsection{Entropy Computing}

Entropy computing is an emerging paradigm being developed by Quantum Computing Inc. Although the currently implemented version is based on measurement and feedback \cite{nguyen2024entropycomputing}, which is not interesting to the present discussion, there are efforts to construct a more quantum version. As with coherent Ising machines, it should be assumed for the purposes of this paper that whenever we discuss entropy computing, we mean an all-optical implementation. Both entropy computing and coherent Ising machines are realized by balancing loss and gain, as discussed in \cite{Berwald2025Zeno}, and the presence of both seems to be important for algorithm performance. Early indications are that this should be a viable path to a quantum advantage, as loss based Zeno effects have recently been shown to support such a speedup \cite{Berwald2024Zeno}. However, the purpose of this paper is not to discuss the entropy computing paradigm in detail.  This will be done in other works. For our purposes, what is important is the encoding used in entropy computing, which is a time-bin encoding, based on the relative number of photons within bins (the current machines work in a regime where this is usually zero or single photons, but we consider general encoding in relative photon number for the present work).

\begin{backmatter}


\bmsection{Acknowledgments}
All authors were completely supported by Quantum Computing Inc.~in the completion of this work. In particular,  NC did not receive any UKRI support. The authors thank Joel Russell Huffman for making aesthetic improvements to the figures.

\bmsection{Disclosures}
All authors hold shares or options in Quantum Computing Inc.

\bmsection{Data availability} All code used in this paper and all data produced are publicly available \cite{dataAvailability}.

\end{backmatter}

\bibliography{references,footnotes}  

\end{document}